%% file: main.tex
\documentclass[11pt,a4paper,english]{article}
\usepackage[utf8]{inputenc} 
\usepackage[T1]{fontenc}    
\usepackage{babel}
\usepackage{blindtext}
\usepackage{hyperref}       
\usepackage{url}            
\usepackage{booktabs}       
\usepackage{amsfonts}       
\usepackage{nicefrac}       
\usepackage{microtype}      
\usepackage{xcolor}         
\usepackage{amsmath,amssymb}
\usepackage{wrapfig}
\usepackage{stfloats}
\usepackage{graphicx}
\usepackage{placeins}  
\usepackage{float}    
\usepackage{caption}
\usepackage{subcaption}
\usepackage{bm}
\usepackage{listings}
\usepackage{multicol}
\usepackage{multirow} 
\usepackage{pgffor}
\usepackage{cleveref}
\usepackage{geometry}

\NewDocumentCommand{\codeword}{v}{%
\texttt{\textcolor{blue}{#1}}%
}

\title{Deep Learning for VWAP Execution in Crypto Markets: \ Beyond the Volume Curve}

\author{%
    Rémi Genet \\
    \small DRM, Université Paris Dauphine - PSL \\
    \small Aplo \\
    \small remi.genet@dauphine.psl.eu \\
}

\begin{document}

\maketitle

\begin{abstract}
Volume-Weighted Average Price (VWAP) is arguably the most prevalent benchmark for trade execution as it provides an unbiased standard for comparing performance across market participants. However, achieving VWAP is inherently challenging due to its dependence on two dynamic factors—volumes and prices. Traditional approaches typically focus on forecasting the market’s volume curve, an assumption that may hold true under steady conditions but becomes suboptimal in more volatile environments or markets such as cryptocurrency where prediction error margins are higher. In this study, I propose a deep learning framework that directly optimizes the VWAP execution objective by bypassing the intermediate step of volume curve prediction. Leveraging automatic differentiation and custom loss functions, my method calibrates order allocation to minimize VWAP slippage, thereby fully addressing the complexities of the execution problem. My results demonstrate that this direct optimization approach consistently achieves lower VWAP slippage compared to conventional methods, even when utilizing a naive linear model presented in \cite{genet2024tln}.They validate the observation that strategies optimized for VWAP performance tend to diverge from accurate volume curve predictions and thus underscore the advantage of directly modeling the execution objective. This research contributes a more efficient and robust framework for VWAP execution in volatile markets, illustrating the potential of deep learning in complex financial systems where direct objective optimization is crucial. Although my empirical analysis focuses on cryptocurrency markets, the underlying principles of the framework are readily applicable to other asset classes such as equities.
\end{abstract}

\newpage

\section{Introduction}\label{section1}
In a rapidly evolving financial landscape, Volume-Weighted Average Price (VWAP) strategies have become a cornerstone for executing high-volume trades, as they aim to minimize market impact and reduce execution costs. VWAP is widely regarded for its fair and neutral calibration, making it an industry standard for comparing performance across market participants \cite{TotalCostOfTransactions}. With the rise of systematic trading, execution strategies have received increasing attention; more specifically, significant institutional trades are now executed using algorithms \cite{Mackenzie}. Despite this, academic research has predominantly focused on Implementation Shortfall (IS) \cite{perold1988implementation} strategies, while VWAP, despite its extensive industrial use and reputation as a robust benchmark (particularly for large orders) \cite{Madhavan2002}, has not been as thoroughly explored. The objective of this study is to propose an alternative VWAP execution approach that moves beyond methods relying solely on forecasting the market volume curve. Indeed, while early approaches in the literature were able to incorporate the price–volume relationship under simplified modeling assumptions, more recent works tend to focus exclusively on the volume curve. This shift arises because more realistic and higher-performing volume models render the simultaneous representation of the volume–volatility (and hence price–volume) relationship considerably more complex. The purpose of this paper is to introduce a method that easily overcomes this limitation, thereby more accurately capturing the full VWAP decision problem. To achieve this, I leverage deep learning as a generalized function calibrator. This approach allows the direct representation of the execution objective in a single step—bypassing the conventional two-step process of first predicting the volume curve and then allocating orders. Traditional econometric and machine learning tools typically rely on this two-step resolution, which can be suboptimal in the presence of prediction errors. The primary challenge in VWAP execution is to optimally allocate trading volumes over the execution horizon—a problem for which historical data does not offer a direct solution. Existing literature often simplifies this challenge by assuming that matching the market's volume curve is ideal. However, this assumption overlooks the fact that perfect volume predictions are unattainable, especially in the volatile cryptocurrency market. Recognizing and incorporating these inevitable prediction errors into the allocation strategy is essential for achieving a price closer to the market VWAP. Moreover, while conventional models can be calibrated to forecast the volume curve from historical data, directly calibrating an execution strategy is not straightforward because the optimal allocation is unknown in advance. Deep learning offers a significant advantage here. Unlike frameworks that minimize the error between predicted and observed data points, deep learning enables the implementation and optimization of any loss function—thanks to automatic differentiation. The central idea of this paper is to build a model that ingests market data, produces an allocation curve for the order, and is trained to minimize the absolute or quadratic error between the achieved execution price and the market VWAP. Finally, although some studies propose dynamic approaches that update predictions throughout an order’s lifetime, I opt for a static framework for two reasons. First, I aim to demonstrate that even a relatively simple network—such as the naive linear model I use—can enhance performance when directly representing the trading strategy. Second, incorporating real-time dynamics introduces additional technological complexities that are beyond the scope of this study. My goal is to propose a simple yet effective method for VWAP execution that fully captures the decision problem and leverages deep learning's flexibility to directly optimize the trader’s true objective.

\section{Related work}

One of the earliest works on VWAP strategies is by \cite{Konishi}, who proposed an optimal slicing strategy for VWAP execution. In scenarios where volume and volatility are uncorrelated, the optimal execution curve aligns with the relative market volume curve; deviations in the correlated case were also quantified. Building on this, \cite{Culoch2007} extended Konishi’s model by incorporating a constrained trading rate and an additional drift term—reflecting situations where traders or brokers, armed with sensitive information, may attempt to “beat” the VWAP. Their study also derived stylized facts regarding the expected relative volume (normalized by trade count), observing an S-shaped pattern throughout the trading day for equities, with higher-turnover stocks exhibiting less variability. These patterns echo the well-known U-shaped trading activity in equity markets. Later, \cite{Culoch2012} advanced this work by developing a continuously dynamic model, demonstrating that an optimal VWAP trading strategy is closely linked to intraday volume estimation. Volumes have also been used as covariates to explain other market variables such as price or volatility \cite{Easley1987, Foster, Tauchen1983, Karpoff1987}, typically using lower-frequency data rather than intraday figures. Additionally, studies such as \cite{Gourieroux} have focused on measuring overall market trading activity through volume metrics.

\medskip

In 2008, \cite{LeFol2006} proposed a method for estimating intraday volumes by decomposing them into two components—an approach refined in \cite{LeFol2012} by separating volumes into a market evolution component and a stock-specific pattern. The dynamic component was modeled using ARMA and SETAR techniques, resulting in significantly improved accuracy over static volume curve approaches. However, while earlier works like \cite{Konishi} and \cite{Culoch2007} attempted to capture the interrelation between price and volume within relatively simple frameworks, \cite{LeFol2006} and \cite{LeFol2012} focused exclusively on the volume component, as incorporating a realistic volume–price relationship renders overall modeling significantly more complex. An alternative execution method is presented in \cite{Humphery}, where Dynamic VWAP (DVWAP) is introduced in contrast to the conventional Historical VWAP (HVWAP). It is argued that incoming news during execution can affect volumes in both directions—information that HVWAP ignores. Their DVWAP framework, built upon earlier methodologies (e.g., those of Bialkowski, Darolles, and Le Fol), not only leverages these improvements but further enhances VWAP performance. Other approaches include stochastic methods proposed by \cite{bouchard} and \cite{frei}. As noted by Frei and Westray, these models derive an optimal trading rate that depends solely on volume curves, neglecting price dynamics due to the assumption that the price process behaves as an uncorrelated Brownian motion. In \cite{Tianhui}, the focus shifts partially to the strategic aspects of VWAP execution at high frequencies, addressing the dilemma faced by brokers between using aggressive versus passive orders. Finally, \cite{Gueant} introduced two novel contributions. First, they integrated both temporary and permanent market impact components—an important consideration for large institutional orders aiming to minimize market impact. Second, they proposed a method for pricing guaranteed VWAP services using a CARA utility function for the broker and indifference pricing. Unlike other approaches that solely strive for benchmark proximity, their framework emphasizes achieving optimal execution while managing risk. A common theme among these approaches is the predominant focus on modeling market volumes, often under the assumption that prices and volumes are independent—a simplification that does not fully capture market reality. Until recently, the latest advancements in machine learning—particularly deep learning—had not been fully leveraged for VWAP execution. Recent studies, however, have begun to explore these novel methods. For instance, \cite{li2022hierarchical} proposed the Macro-Meta-Micro Trader (M3T) architecture, which combines deep learning with hierarchical reinforcement learning to capture market patterns and execute orders across multiple temporal scales. Although this approach achieved an average cost saving of 1.16 basis points relative to an optimal baseline, its complexity may hinder practical implementation. Similarly, \cite{kim2023adaptive} developed a dual-level reinforcement learning strategy using Proximal Policy Optimization (PPO) to track daily cumulative VWAP. Their method integrates a Transformer model to capture the overall U-shaped volume pattern with an LSTM model for finer order distribution. Despite improved accuracy over previous reinforcement learning models, it remains heavily reliant on volume prediction. In \cite{papanicolaou2023optimal}, large stock order execution is simulated using LSTMs within the Almgren and Chriss framework. By leveraging cross-sectional data from multiple stocks to capture inter-stock dependencies, their approach consistently outperforms TWAP and VWAP-based strategies on S\&P100 data. However, the focus remains on minimizing transaction costs rather than directly optimizing VWAP execution. While recent studies underscore the potential of machine learning techniques for VWAP execution, many of these methods continue to depend on traditional volume curve prediction rather than directly targeting the VWAP execution objective.

\section{Defining the VWAP Problem}

\medskip

The \emph{Volume Weighted Average Price (VWAP)} is a trading benchmark that reflects the average price at which a security is traded throughout the day, weighted by volume. It provides insights into both the trend and the value of a security.
\begin{itemize}
  \item Its basic formulation is:
  \begin{equation}
    \text{VWAP} = \frac{\text{Total Traded Value}}{\text{Total Traded Volume}}.
  \end{equation}
  \item More explicitly, VWAP is computed as:
  \begin{equation}
    \text{VWAP} = \frac{\sum_{i} \left(p_i \times v_i\right)}{\sum_{i} v_i},
  \end{equation}
where \( p_i \) denotes the price of the \(i\)th trade and \( v_i \) the corresponding volume.
\end{itemize}

\subsection*{Defining the Execution Problem}
The objective is to execute trades so that the average execution price aligns as closely as possible with the market VWAP over a given period \(T\). The total order volume is denoted by \(v\) and the total market volume over the same period by \(V\).

The average execution price is defined as:
\begin{equation}
    P = \frac{\sum_{j=1}^{n}{ p_j \times v_j}}{v}, \quad \text{with } v = \sum_{j=1}^{n}{v_j},
\end{equation}
assuming that \(n\) transactions are executed during period \(T\).

Similarly, the market VWAP is defined as:
\begin{equation}
    \text{VWAP} = \frac{\sum_{i=1}^{N}{ p_i \times V_i}}{V}, \quad \text{with } V = \sum_{i=1}^{N}{V_i},
\end{equation}
where \(N\) represents the total number of market transactions (including the order's own trades).

\subsection*{Addressing Slippage}
Absolute slippage is defined as:
\begin{equation}
    S = \left| P - \text{VWAP} \right|.
\end{equation}
Minimizing \(S\) is challenging because the transaction space is non-standard—trades occur as discrete events at irregular intervals rather than in uniform time or volume increments. To simplify the analysis, the overall period is divided into \(T\) smaller intervals. Within each interval, the following definitions apply:
\begin{equation}
    \text{VWAP}_{T} = \frac{\sum_{t=1}^{T}{ \text{VWAP}_t \times V_t}}{V}, \quad P_{T} = \frac{\sum_{t=1}^{T}{ P_t \times v_t}}{v},
\end{equation}
where:
\begin{itemize}
  \item \(V_t\) is the market volume in interval \(t\),
  \item \(v_t\) is the executed volume in interval \(t\),
  \item \(\text{VWAP}_t\) and \(P_t\) denote the market VWAP and the average execution price in interval \(t\), respectively.
\end{itemize}
This segmentation permits a more detailed comparison between \(P_T\) and \(\text{VWAP}_T\).

\subsection*{Refining the Problem Definition}
For further analysis, normalized volume proportions are introduced:
\[
Q_t = \frac{V_t}{V} \quad \text{and} \quad q_t = \frac{v_t}{v},
\]
so that:
\[
\sum_{t=1}^{T} Q_t = \sum_{t=1}^{T} q_t = 1.
\]
Thus, the execution price and market VWAP can be rewritten as:
\[
P_T = \sum_{t=1}^{T} P_t \, q_t \quad \text{and} \quad \text{VWAP}_T = \sum_{t=1}^{T} \text{VWAP}_t \, Q_t.
\]
Accordingly, the slippage becomes:
\begin{equation} \label{eq:static_slippage1}
    S_T = \left| P_T - \text{VWAP}_T \right| = \left| \sum_{t=1}^{T} \left( P_t \, q_t - \text{VWAP}_t \, Q_t \right) \right|.
\end{equation}
A common strategy to decompose the difference is to add and subtract a common term inside the summation. In this case, \(\text{VWAP}_t \, q_t\) is added and subtracted for each interval:
\begin{equation}
P_t \, q_t - \text{VWAP}_t \, Q_t = \underbrace{\left(P_t \, q_t - \text{VWAP}_t \, q_t\right)}_{=(P_t-\text{VWAP}_t)q_t} + \underbrace{\left(\text{VWAP}_t \, q_t - \text{VWAP}_t \, Q_t\right)}_{=\text{VWAP}_t\,(q_t-Q_t)}.
\end{equation}
Substituting this into \eqref{eq:static_slippage1} yields:
\begin{equation}
    S_T = \left| \sum_{t=1}^{T} \left[(P_t-\text{VWAP}_t)q_t + \text{VWAP}_t\,(q_t-Q_t)\right] \right|.
\end{equation}
Since the absolute value of a sum is generally not equal to the sum of the absolute values, the triangle inequality is invoked:
\[
\left|\sum_{t=1}^{T} a_t\right| \le \sum_{t=1}^{T} \left|a_t\right|.
\]
Thus,
\begin{equation} \label{eq:static_slippage-bound}
    S_T \le \sum_{t=1}^{T} \left| (P_t-\text{VWAP}_t)q_t + \text{VWAP}_t\,(q_t-Q_t) \right|.
\end{equation}
Furthermore, applying the triangle inequality to each individual term results in:
\begin{equation}
    \begin{aligned}
    &\left| (P_t-\text{VWAP}_t)q_t + \text{VWAP}_t\,(q_t-Q_t) \right| \\
    &\qquad \leq \left| (P_t-\text{VWAP}_t)q_t \right| + \left| \text{VWAP}_t\,(q_t-Q_t) \right|.
    \end{aligned}
\end{equation}
Summing over \(t\) gives:
\begin{equation}
    S_T \le \sum_{t=1}^{T} \left| (P_t-\text{VWAP}_t)q_t \right| + \sum_{t=1}^{T} \left| \text{VWAP}_t\,(q_t-Q_t) \right|.
\end{equation}
In this decomposition:
\begin{itemize}
    \item \(\left| (P_t-\text{VWAP}_t)q_t \right|\) represents the deviation in price weighted by the executed volume proportion.
    \item \(\left| \text{VWAP}_t\,(q_t-Q_t) \right|\) captures the impact of discrepancies between the executed volume allocation and the market’s volume distribution.
\end{itemize}

\subsection*{Methodology --- Delineating the Problems}
In the pursuit of minimizing slippage, two distinct challenges are identified.

\subsubsection*{Problem 1: Price Deviation Minimization}
This aspect focuses on minimizing:
\[
\sum_{t=1}^{T} \left| (P_t-\text{VWAP}_t)q_t \right|.
\]
Key considerations include:
\begin{itemize}
    \item Execution quality within each interval is paramount; the goal is to achieve prices as close as possible to the market VWAP.
    \item With finer time intervals, price deviations are expected to be smaller, and deviations across bins may partially offset one another.
    \item This challenge is primarily related to market microstructure and the quality of executions.
\end{itemize}

\subsubsection*{Problem 2: Volume Allocation Optimization}
This challenge is characterized by:
\begin{equation} \label{eq:static_volumeOptimization}
    \sum_{t=1}^{T} \left| \text{VWAP}_t\,(q_t-Q_t) \right|.
\end{equation}
Key considerations include:
\begin{itemize}
    \item Although \(\text{VWAP}_t\) cannot be controlled, the executed volume allocation \(q_t\) can be adjusted.
    \item Traditional methods focus on accurately predicting \(Q_t\), the market volume profile; however, given the inherent uncertainty in market dynamics, this may not be optimal.
    \item The strategy is to design a \(q_t\) allocation that accounts for the noisy nature of the \(Q_t\) process, adapting to market conditions and prediction errors.
    \item For instance, during periods of low volatility, it may be advisable to execute less than the market proportion \(Q_t\) to mitigate the risk of sudden volatility spikes.
    \item The aim is to develop a robust volume allocation strategy that minimizes slippage even when predictions of \(Q_t\) are imperfect.
\end{itemize}

\subsubsection*{Choosing the Appropriate Focus}
While the primary aim is to minimize absolute slippage, it is also essential to consider strategies that not only meet the benchmark but also potentially outperform it.

\subsubsection*{Price Difference Focus}
\begin{itemize}
    \item Since the weights \(q_t\) and \(Q_t\) are non-negative, achieving the most favorable prices in each interval is crucial.
    \item Opportunities for value addition may arise, particularly through the mitigation of trading costs such as the spread.
\end{itemize}

\subsubsection*{Volume Allocation Focus}
\begin{itemize}
    \item Given the challenges in precisely forecasting \(\text{VWAP}_t\), emphasis is placed on risk reduction through effective volume allocation.
    \item The strategy aims to minimize the term \(\left| \text{VWAP}_t\,(q_t-Q_t) \right|\) by aligning the executed volume with the market’s volume profile.
    \item This involves leveraging insights into future market volumes and understanding their interaction with market volatility.
    \item The focus is on developing the optimal \(q_t\) strategy for volume allocation over larger time intervals, thereby reducing the risk associated with VWAP execution.
\end{itemize}

\medskip

In summary, by rigorously decomposing the slippage \(S_T\) and carefully applying the triangle inequality, the problem is separated into two distinct components: execution quality (price differences) and volume allocation discrepancies. This detailed derivation provides a solid foundation for subsequent analysis and strategy development.

\section{A fixed optimal allocation curve approach}\label{section3}
To demonstrate the validity of allocation strategies that are not based solely on following the volume curve, an analysis was conducted in which an optimal allocation curve was calibrated using only historical data, without incorporating current market information. In this experiment, the allocation curve is represented by a simple vector of weights that determines the fraction of the total order executed in each time bin. Without any additional market inputs, the expected volume curve is a uniform distribution over the execution period—precisely as conventional wisdom would predict—so any deviation from this flat pattern in the optimal allocation would challenge that notion.

\subsection{Methodology}

\begin{figure}[!htb]
    \centering
    \includegraphics[width=0.9\textwidth]{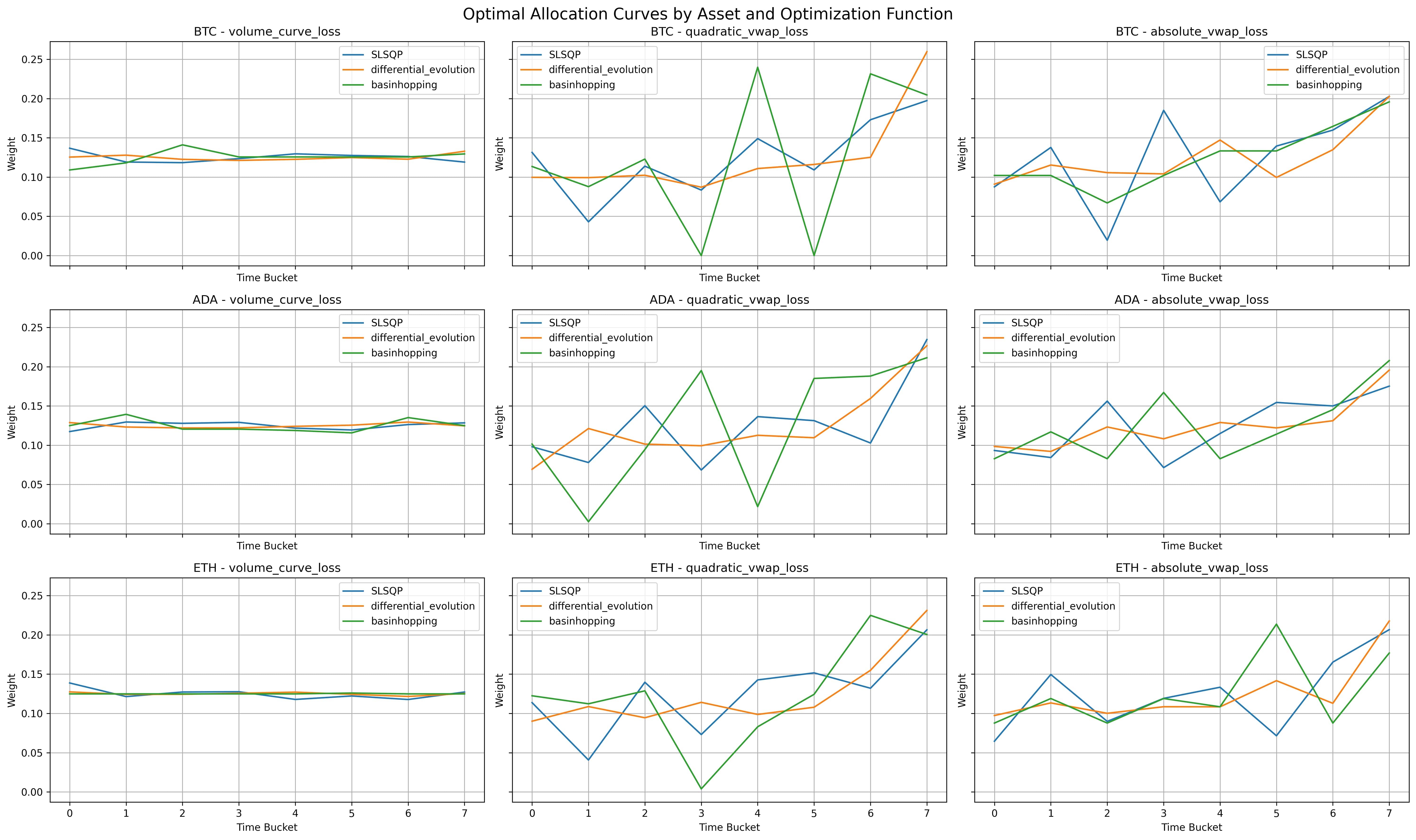}
    \caption{Optimal Allocation based on optimization methods and asset}
    \label{fig:static_fixed_optimal_allocation}
\end{figure}

Historical data from Binance perpetual futures contracts for Bitcoin (BTC), Ethereum (ETH), and Cardano (ADA) covering the period from contract inception until July 1, 2024, were employed. The data, recorded at an hourly frequency, were divided into training and testing sets in an 80/20 split, with the temporal order preserved to ensure a realistic evaluation. The VWAP execution problem was formulated as the optimization of a vector of allocation weights, \(q_t\) (for \(t = 1, \dots, T\) with \(T=8\) bins), by minimizing one of three loss functions. The first is the Quadratic VWAP Loss, defined as
\begin{equation} \label{eq:static_quad_loss}
L_Q = \mathbb{E}\!\left[\left(\frac{\text{VWAP}_{\text{achieved}}}{\text{VWAP}_{\text{market}}} - 1\right)^2\right],
\end{equation}
the second is the Absolute VWAP Loss,
\begin{equation} \label{eq:static_abs_loss}
L_A = \mathbb{E}\!\left[\left|\frac{\text{VWAP}_{\text{achieved}}}{\text{VWAP}_{\text{market}}} - 1\right|\right],
\end{equation}
and the third is the Volume Curve Loss,
\begin{equation} \label{eq:static_vol_curve_loss}
L_V = \mathbb{E}\!\left[\sum_{t=1}^{T} \left(\frac{v_t}{\sum_{s=1}^{T} v_s} - \frac{V_t}{\sum_{s=1}^{T} V_s}\right)^2\right].
\end{equation}
The achieved and market VWAP are defined respectively as
\begin{align}
\text{VWAP}_{\text{achieved}} &= \frac{\sum_{t=1}^{T} v_t\, p_t}{\sum_{t=1}^{T} v_t}, \label{eq:static_achieved_vwap} \\
\text{VWAP}_{\text{market}} &= \frac{\sum_{t=1}^{T} V_t\, p_t}{\sum_{t=1}^{T} V_t}. \label{eq:static_market_vwap}
\end{align}
Here, \(v_t\) represents the allocated volume in bin \(t\), \(V_t\) is the market volume in bin \(t\), and \(p_t\) is the VWAP price in bin \(t\). To solve the optimization problem, three different methods were employed. Sequential Least Squares Programming (SLSQP) was used as a local optimization technique with 1000 restarts from random initial guesses to mitigate the risk of converging to local minima. Basin-Hopping, a global optimization algorithm that combines local searches with random perturbations, was also employed and performed 10 times. In addition, Differential Evolution, an evolutionary algorithm known for its robustness in finding global optima, was executed once. Performance metrics for each asset, loss function, and optimization method are summarized in Table~\ref{tab:static_fix_curve_vwap_results}, while Figure~\ref{fig:static_fixed_optimal_allocation} illustrates the optimal allocation curves derived from these optimizations.

\input{tables/2_fix_curve_table}

The analysis reveals several important findings. Allocation curves obtained by minimizing the absolute and quadratic VWAP losses clearly deviate from a uniform allocation, while the optimization based on the volume curve loss yields a nearly flat allocation. This flat allocation is expected since, in the absence of market information, the average volume in each time bin is equal. However, although the flat allocation perfectly tracks the average volume curve (as evidenced by an \(R^2\) near zero), it performs worse in terms of VWAP execution compared to the non-uniform allocations derived from the VWAP-based losses. Furthermore, among the assets studied, Bitcoin consistently exhibits the lowest VWAP losses, reflecting its relative stability, whereas Cardano—being more volatile—shows higher losses. In terms of optimization methods, Differential Evolution consistently produced competitive results across assets and loss functions, while SLSQP, despite numerous restarts, often converged to suboptimal solutions due to the complexity of the optimization landscape. Basin-Hopping displayed variable performance that was generally less consistent.

\subsection{Discussion}

This experiment is particularly revealing because it involves optimizing a vector of allocation weights using only historical data without current market information. Consequently, an optimization based solely on the volume curve loss produces a flat, uniform allocation, which is expected when no additional information beyond the historical average is available. In contrast, the allocation curves optimized using the absolute and quadratic VWAP losses diverge significantly from uniformity, exhibiting a pronounced end-loading behavior in some cases. This divergence indicates that even in the absence of direct market signals, an allocation strategy that minimizes VWAP slippage is inherently non-uniform. This finding challenges the traditional approach of following the expected volume curve and strongly supports the use of direct VWAP loss optimization in trading execution strategies. Although the empirically derived allocation curves deviate markedly from the conventional volume curve, this outcome is not entirely surprising when considering simplified frameworks that account for price--volume interactions. For instance, Konishi \cite{Konishi} shows that the optimal fraction \(x^*(t)\) allocated to bin \(t\) can be approximated by
\begin{equation}\label{eq:static_konishi_approx}
x^*(t) \;\approx\; \frac{\mathbb{E}\bigl[\sigma(t,V)^2 \, X(t)\bigr]}{\mathbb{E}\bigl[\sigma(t,V)^2\bigr]},
\end{equation}
or equivalently, as rewritten by McCulloch and Kazakov \cite{Culoch2007}:
\begin{equation}\label{eq:static_cullocheq}
x_t 
= \frac{\mathbb{E}[X_t \sigma_t^2]}{\mathbb{E}[\sigma_t^2]} 
= \mathbb{E}[X_t] 
+ \frac{\mathrm{Cov}[X_t, \sigma_t^2]}{\mathbb{E}[\sigma_t^2]}.
\end{equation}
Thus, a sufficiently large positive covariance term can shift the solution away from the market's average volume curve, underscoring the role of volatility in shaping optimal allocations. These results highlight the importance of the price--volume relationship and expose the limitations of approaches based solely on replicating the historical volume curve. Although more complex models can be developed to jointly account for both price and volume dynamics, the present analysis demonstrates that even a relatively simple (and naive) linear model can yield meaningful improvements in VWAP execution by directly optimizing the VWAP objective.

\section{A Deep-Learning Model for VWAP Allocation}
\subsection{Framework}

Having demonstrated that the traditional volume curve approach may not be optimal for allocating volumes over time, deep learning is employed as a powerful means to represent and optimize the VWAP allocation problem. Deep learning affords the flexibility to directly optimize the specific objective—minimizing the deviation between the achieved execution price and the market VWAP—by leveraging automatic differentiation. In this framework, the model accepts sequential inputs comprising multiple features over a designated historical lookback period and produces an allocation curve for future time intervals. The key innovation is the use of a custom loss function (either absolute or quadratic VWAP loss) that captures the true execution error, thereby enabling the optimization of the strategy in a single step rather than relying on intermediate proxies such as the volume curve. A critical requirement is that the model’s output constitutes a valid allocation strategy; that is, the allocation weights must be non-negative and sum to one. Although several approaches (e.g., clipping negative values or taking absolute values before normalization) were considered, these methods risk introducing uneven optimization. Instead, the softmax function is adopted, which naturally transforms any real-valued vector into a smooth probability distribution over the execution horizon. This ensures that even if the Temporal Linear Network (TLN) produces negative values, the final allocation vector remains both non-negative and normalized, thereby satisfying the constraints without sacrificing optimization smoothness.

\subsection{Internal Model}

In selecting an internal model for the VWAP allocation framework, various architectures traditionally employed for sequential prediction tasks were considered. Recent studies in volume prediction for cryptocurrency markets have explored models such as Temporal Kolmogorov-Arnold Networks (TKAN) \cite{genet2024tkan}, Signature-Weighted Kolmogorov-Arnold Networks (SigKAN) \cite{inzirillo2024sigkan}, Temporal Kolmogorov-Arnold Transformers (TKAT) \cite{genet2024tkat}, Kolmogorov-Arnold Mixture of Experts (KAMoE) \cite{inzirillo2024kamoe} and Recurrent Neural Networks with Signature-Based Gating Mechanisms (SigGate) \cite{genet2025siggate}. Although these approaches offer innovative solutions for capturing long-term dependencies and complex interactions, they also introduce additional complexity that is not necessary when the primary goal is to demonstrate the importance of optimizing the correct objective. Accordingly, the Temporal Linear Network (TLN) \cite{genet2024tln} was chosen as the default internal model. The TLN is conceptually simple and highly interpretable—essentially a linear model with structured transformations—yet it is integrated within a deep learning framework that enables the minimization of arbitrary loss functions via automatic differentiation. Importantly, even a naive linear model performs effectively within the proposed approach, emphasizing that improvements arise from directly optimizing the VWAP execution objective rather than from architectural complexity.
Let the input tensor be
$$
X \in \mathbb{R}^{B \times S \times F},
$$
where \(B\) is the batch size, \(S\) is the length of the input sequence, and \(F\) is the number of input features. The TLN consists of \(L\) layers, each implementing a cascade of linear operations. Each operation within a layer is a linear mapping with learnable weights, and the layers are designed to progressively adjust both the sequence length and feature dimension. This structured design leverages parameter sharing and constraints, resulting in a model that is both computationally efficient and stable.

Each layer \(\ell\) (with \(1 \leq \ell \leq L\)) transforms its input \(X^{(\ell-1)}\) into an output \(X^{(\ell)}\) through two primary stages: a \textbf{structured linear transformation} and a \textbf{convolutional filtering} along the time dimension.

\subsubsection*{1. Structured Linear Transformation}

This stage is composed of three consecutive linear operations.

\paragraph{Temporal Scaling:}  
Each time index is scaled by a learnable vector \(K^{(\ell)} \in \mathbb{R}^{S_\ell}\), where \(S_\ell\) denotes the current sequence length at layer \(\ell\). For each batch index \(b\), time index \(s\), and feature index \(f\), the computation is as follows:
\begin{equation}
    \widetilde{X}^{(\ell)}_{b,s,f} = K^{(\ell)}_s \, X^{(\ell-1)}_{b,s,f}.
\end{equation}

\paragraph{Feature Transformation:}  
At each time step, the features are mapped from \(\mathbb{R}^{F_\ell}\) to an intermediate space \(\mathbb{R}^{F^\prime_\ell}\) using a learnable weight matrix \(W^{(\ell)} \in \mathbb{R}^{F_\ell \times F^\prime_\ell}\) and bias \(b^{(\ell)} \in \mathbb{R}^{F^\prime_\ell}\):
\begin{equation}
    Y^{(\ell)}_{b,s,\cdot} = \widetilde{X}^{(\ell)}_{b,s,\cdot}\, W^{(\ell)} + b^{(\ell)}.
\end{equation}
An additional element-wise scaling is then applied using a learnable vector \(F^{(\ell)} \in \mathbb{R}^{F^\prime_\ell}\):
\begin{equation}
    Z^{(\ell)}_{b,s,f^\prime} = F^{(\ell)}_{f^\prime} \cdot Y^{(\ell)}_{b,s,f^\prime}.
\end{equation}

\paragraph{Temporal Transformation:}  
Finally, the sequence is remapped to a new length \(S^\prime_\ell\) by applying a linear transformation with learnable parameters \(T^{(\ell)} \in \mathbb{R}^{S_\ell \times S^\prime_\ell}\) and bias \(c^{(\ell)} \in \mathbb{R}^{S^\prime_\ell}\):
\begin{equation}
    \widehat{Z}^{(\ell)}_{b,s^\prime} = \sum_{s=1}^{S_\ell} T^{(\ell)}_{s,s^\prime}\, Z^{(\ell)}_{b,s,f^\prime} + c^{(\ell)}_{s^\prime}.
\end{equation}
In summary, the structured transformation is represented as
\begin{equation}
    \Phi^{(\ell)}\bigl(X^{(\ell-1)}\bigr) = \mathcal{T}^{(\ell)}\!\left( \left( X^{(\ell-1)} \odot K^{(\ell)} \right)W^{(\ell)} + b^{(\ell)} \odot F^{(\ell)} \right),
\end{equation}
where \(\mathcal{T}^{(\ell)}(\cdot)\) denotes the temporal mapping and \(\odot\) represents element-wise multiplication.

\subsubsection*{2. Convolutional Filtering}

After the structured transformation, each feature channel is convolved along the time dimension. Let \(C^{(\ell)} \in \mathbb{R}^{K^{(\ell)}_{\text{conv}} \times F^\prime_\ell}\) denote the convolution kernel (with kernel size \(K^{(\ell)}_{\text{conv}}\)) and \(d^{(\ell)} \in \mathbb{R}^{F^\prime_\ell}\) the corresponding bias. The convolution is computed as
\begin{equation}
    X^{(\ell)}_{b,t,f^\prime} = \sum_{k=0}^{K^{(\ell)}_{\text{conv}}-1} C^{(\ell)}_{k,f^\prime} \, \Phi^{(\ell)}\bigl(X^{(\ell-1)}\bigr)_{b,t+k,f^\prime} + d^{(\ell)}_{f^\prime}.
\end{equation}
Thus, the overall operation in layer \(\ell\) is given by
\begin{equation}
    X^{(\ell)} = \operatorname{Conv}\!\Bigl( \Phi^{(\ell)}\bigl(X^{(\ell-1)}\bigr) \Bigr).
\end{equation}

\subsection{The Role of Softmax in Enforcing Allocation Constraints}

A key innovation in the framework is the use of the softmax function to enforce allocation constraints. In this context, the final output of the internal model must be a valid allocation vector—that is, all elements must be non-negative and the vector must sum to one. While straightforward normalization (dividing by the sum of the outputs) or clipping negative values might appear sufficient, these approaches can lead to uneven or suboptimal optimization. In contrast, the softmax function
\begin{equation}
    \sigma(z)_i = \frac{e^{z_i}}{\sum_{j=1}^K e^{z_j}},
\end{equation}
naturally transforms any real-valued input vector \(z\) into a smooth probability distribution. This not only satisfies the constraints but also ensures a differentiable transformation that facilitates gradient-based optimization. Even if the TLN produces negative values, the exponential function within softmax guarantees a smooth and balanced allocation across the execution horizon.

\subsection{Comparison with Standard Linear Regression and Rationale}

A conventional linear regression would flatten the input tensor \(X \in \mathbb{R}^{B \times S \times F}\) into \(X_{\text{flat}} \in \mathbb{R}^{B \times (S \cdot F)}\) and compute 
\begin{equation}
    Y = X_{\text{flat}}\, W_{\text{reg}} + b_{\text{reg}},
\end{equation}
resulting in a very large number of parameters. In contrast, the TLN architecture leverages structured, layer-by-layer transformations to progressively adjust the sequence and feature dimensions, yielding a much more parsimonious model that is less prone to unstable weight estimates. Furthermore, by embedding the TLN within a deep learning framework, it becomes possible to optimize non-traditional loss functions—such as the VWAP-specific losses—using automatic differentiation, an advantage that standard linear regression does not offer.

\subsection{Summary of the Optimization Process}
The overall optimization process proceeds in three primary steps. First, the TLN processes the input features over a specified lookback period to generate a raw output vector \(v_{[t, t+h]} = \text{TLN}(x_t)\) for future time steps. Next, the softmax function is applied to these raw outputs to convert them into a valid allocation vector:
\begin{equation}
    q_t = \frac{e^{v_t}}{\sum_{s=1}^{T} e^{v_s}}, \quad \text{for } t = 1, 2, \dots, T,
\end{equation}
ensuring that the allocations are non-negative and sum to one. Finally, the model parameters \(\theta\) are optimized by minimizing the quadratic difference between the achieved VWAP and the market VWAP, as defined in Section~\ref{section3} (see Eqs.~\eqref{eq:static_quad_loss}--\eqref{eq:static_market_vwap}). In summary, this framework establishes a fair comparison platform by leveraging deep learning to directly optimize the VWAP execution objective. Rather than increasing model complexity, the approach employs a straightforward, linear model (the TLN) enhanced by the softmax transformation to enforce smooth and constrained allocation, demonstrating that aligning the optimization objective precisely with the trading goal can lead to improved execution performance.

\section{Results}
In this section, a comprehensive comparison of various VWAP execution strategies is presented, including traditional methods, the proposed deep-learning approach, and dynamic strategies. These methods are evaluated across different cryptocurrencies to assess their performance under varying market conditions.

\subsection{Benchmarks and Strategies}

Several benchmark strategies are considered. As a baseline, a naive flat allocation strategy that uniformly distributes the order volume across all time intervals is included; this simple method serves as a reference for assessing more sophisticated approaches. In addition, fixed optimal allocation curves are evaluated—static allocations optimized using three different loss functions (absolute VWAP loss, quadratic VWAP loss, and volume curve loss) as described in \ref{section3}. For this optimization, the differential evolution algorithm is employed due to its demonstrated superior and more stable performance. The proposed deep-learning approach, referred to as the \emph{StaticVWAP Model}, employs a neural network to generate allocation curves based on historical market data. For a fair comparison, all models receive the same input features, including volumes, the hour of the day, the day of the week, and returns computed on the VWAP price of each bin. In contrast, the \emph{Static Linear Regression} model serves as a traditional benchmark that is calibrated to predict the volume. Since linear regression does not inherently produce outputs that satisfy the allocation constraints (i.e., non-negativity and summing to one), its predictions are post-processed by first clipping negative values to zero and then normalizing the resulting vector. In cases where all values are zero (or become zero after clipping), an equiponderated allocation is returned. Dynamic VWAP strategies based on the framework introduced by \cite{LeFol2012} are also implemented. In these dynamic strategies, the volume executed at each time step is determined by the current market volume prediction and the remaining order size, following the equation
\begin{equation}
    v_t = \frac{\hat{V}_t}{\sum_{i=t}^T \hat{V}_i} \cdot \left(1 - \sum_{i=0}^{t-1} v_i\right),
\end{equation}
where \(v_t\) denotes the volume executed at time \(t\), \(\hat{V}_t\) is the predicted market volume at time \(t\), \(T\) is the end of the execution horizon, and \(\sum_{i=0}^{t-1} v_i\) represents the cumulative executed volume up to time \(t-1\). Both a dynamic linear regression approach and a dynamic version of the StaticVWAP model, which updates predictions at each time step, are considered.

\subsection{Evaluation Metrics}

Three key metrics are used to assess performance. The \emph{Absolute VWAP Loss} measures the absolute difference between the achieved VWAP and the market VWAP, while the \emph{Quadratic VWAP Loss} measures the squared difference, thereby penalizing larger deviations more heavily. In addition, the \emph{\(R^2\) Score for the Volume Curve} is computed, indicating how accurately the strategy predicts the actual market volume profile.

\subsection{Experimental Setup}

The strategies are evaluated using hourly data from five major cryptocurrencies (BTC, ETH, BNB, ADA, and XRP) traded on Binance perpetual contracts covering the period from January 1, 2020, to July 1, 2024. This dataset provides sufficient historical data across varying market conditions to evaluate the models' performance in both high and low volatility regimes. The data is partitioned chronologically, with the last 20\% reserved as the test set to ensure a true out-of-sample evaluation. A validation set, comprising 20\% of the remaining data, is randomly selected from the first 80\% of the period to guide model selection and prevent overfitting to any specific market regime.

For input features, several categories are incorporated: (1) Volume data, which is transformed to address non-stationarity by dividing each volume value by a trailing two-week moving average, calculated with appropriate shifting to avoid look-ahead bias; (2) Temporal indicators including hour of day and day of week as categorical features to capture seasonality patterns; and (3) Price information in the form of returns calculated on the VWAP price of each bin, with a value of 0 assigned when no volume is traded.

The preparation of target variables differs based on the model type. For the neural network approaches, the normalization to volume curve is implemented directly within the loss function, with volumes divided by the sum over the lookback period for numerical stability. For traditional linear regression models, the volume curve is explicitly normalized by dividing each volume by the sum of volumes in the lookahead period, as these models cannot incorporate this normalization within their loss functions.

In these experiments, a lookback period of 120 time steps is used to predict allocations for 12 periods ahead. Additional experiments with modified parameters—one with a 6-step prediction horizon and a shorter lookback period, and another with a 48-step prediction horizon—are provided in the Appendix. The results from these experiments confirm that the framework is effective across different timeframes and that the underlying concepts generalize well.

\subsection{Training Details}

Standard training settings were adopted. Specifically, the Adam optimizer was employed, and the maximum number of epochs was set to 1000. To mitigate overfitting and ensure stable convergence, two key callbacks were incorporated: an early stopping mechanism, which halts training if no improvement in validation loss is observed for 10 consecutive epochs, and a learning rate reducer, which divides the current learning rate by 4 after 5 epochs without improvement (starting from an initial rate of 0.001). The training set was shuffled, and a 20\% validation split was used to monitor these metrics during training.

An additional critical aspect was the mitigation of sensitivity to weight initialization. Each training run was repeated 30 times with different random initializations to compute both the average performance and its standard deviation, thereby providing a robust and statistically sound evaluation of model effectiveness.

\subsection{Empirical Results}

Table~\ref{tab:static_vwap_results} summarizes the performance of various VWAP execution strategies for a 12-step-ahead allocation with a 120-step lookback window. The table reports the mean and standard deviation of the Absolute VWAP Loss (scaled by \(10^{-2}\)), Quadratic VWAP Loss (scaled by \(10^{-4}\)), the \(R^2\) score for the volume curve, and the training time (in seconds) for each strategy, across five assets (BTC, ETH, BNB, ADA, and XRP).

Several key patterns emerge from these results. First, all proposed methods consistently outperform the naive flat allocation approach across all assets, confirming that more sophisticated allocation strategies yield tangible benefits for VWAP execution. 

\input{tables/5_results_table}

When comparing the methods based on their ability to predict volume curves, we observe that Static Linear Regression is more effective than the StaticVWAP model when the latter is optimized for absolute or quadratic VWAP loss. However, the Dynamic Linear Regression approach consistently achieves the highest \(R^2\) values across all assets, indicating superior volume curve prediction capability. This improved prediction translates into better VWAP execution compared to Static Linear Regression.

\medskip

Interestingly, despite Dynamic Linear Regression's advantage in volume prediction, the Fixed Volume Curve approach demonstrates superior VWAP execution performance. This finding underscores a critical insight: accurate volume curve prediction, while beneficial, is not the primary determinant of optimal VWAP execution.

\medskip

Most significantly, the proposed StaticVWAP deep learning model consistently achieves the best absolute or quadratic VWAP loss values (depending on its optimization target) across all assets. When trained to minimize absolute VWAP loss, it delivers the lowest absolute slippage; similarly, when optimized for quadratic VWAP loss, it produces the lowest quadratic slippage metrics.

\medskip

Regarding training times, the deep-learning models (StaticVWAP) generally require between 15 and 18 seconds per run, which, although longer than the near-instantaneous training times of linear regression models, represents an acceptable trade-off given the significant performance gains. The dynamic approaches show mixed results when applied to the StaticVWAP model, often degrading performance rather than improving it.

\medskip

Overall, these results validate the central thesis of this study: directly optimizing the VWAP execution objective—rather than focusing solely on volume curve prediction—is essential for achieving superior execution performance in volatile markets such as cryptocurrencies. The StaticVWAP deep learning framework provides a flexible and effective approach that consistently outperforms conventional methods, even when utilizing a relatively simple model architecture.

\subsection{Training Times and Comparison Across Prediction Horizons}

Tables~\ref{tab:static_vwap_results_6} and~\ref{tab:static_vwap_results_48} provide additional insight into the practical applicability of each method under different prediction horizons. Several consistent patterns emerge across these timeframes that reinforce the findings from our main experiments.

\medskip

First, regarding training efficiency, the StaticVWAP models consistently maintain reasonable computational demands across all horizons—typically under 20 seconds—making them practical for real-world applications despite being more complex than linear regression models. Notably, while the fixed volume curve approach converges rapidly for shorter horizons, its training time increases dramatically with longer prediction horizons, eventually requiring several minutes per run for 48-step predictions.

\medskip

When examining performance across different timeframes, the relative rankings of the models remain remarkably consistent. For both shorter (6-step) and longer (48-step) prediction horizons, the StaticVWAP models optimized directly for VWAP loss consistently outperform both naive approaches and linear regression variants. The fixed volume curve approach remains competitive, particularly when optimized with quadratic loss, but generally does not reach the performance levels of the StaticVWAP model.

\medskip

Dynamic approaches show consistent patterns across different horizons. Dynamic linear regression demonstrates increasingly strong volume curve prediction capability (higher \(R^2\) values) as the prediction horizon extends, yet this improvement in volume prediction does not translate to proportional gains in VWAP execution performance. This pattern strongly supports our central thesis that accurate volume prediction, while informative, is not the primary driver of optimal VWAP execution. In fact, the performance gap between methods directly optimized for VWAP execution and those focused on volume prediction tends to persist or even widen with longer horizons.

\medskip

The stability of these findings across different prediction horizons and assets underscores the robustness of our approach. Whether executing shorter or longer-term VWAP strategies, directly optimizing the execution objective consistently yields superior results compared to approaches that rely primarily on volume curve prediction.

\subsection{Limitations and Future Directions}

Although the proposed approach yields promising improvements in VWAP execution, several limitations warrant discussion. One key assumption in the current framework is that market impact is negligible. This assumption is common in the VWAP literature, as employing a VWAP benchmark over a long period is believed to mitigate market impact. However, it is recognized that large order sizes may affect execution quality. Studies such as Gueant et al. have shown that both temporary and permanent market impact components can be incorporated into the optimization process, suggesting that future work could integrate such models to more accurately capture real-world trading dynamics. Another limitation is that the present model operates in a static framework without dynamic updating in real time. Although static models have the advantage of simplicity and reduced technological complexity, dynamic adjustments could potentially enhance execution performance in highly volatile environments. Future research may explore hybrid approaches that combine the benefits of static optimization with dynamic market adaptation. Finally, while the analysis in this study focuses on cryptocurrency markets, the underlying principles are readily extendable to other asset classes such as equities. Incorporating more sophisticated market impact models and dynamic updating mechanisms will be valuable in developing a comprehensive framework for VWAP execution across diverse trading environments.

\subsection{Understanding the Predicted Curves}

\begin{figure}[H]
    \centering
    \includegraphics[width=\columnwidth]{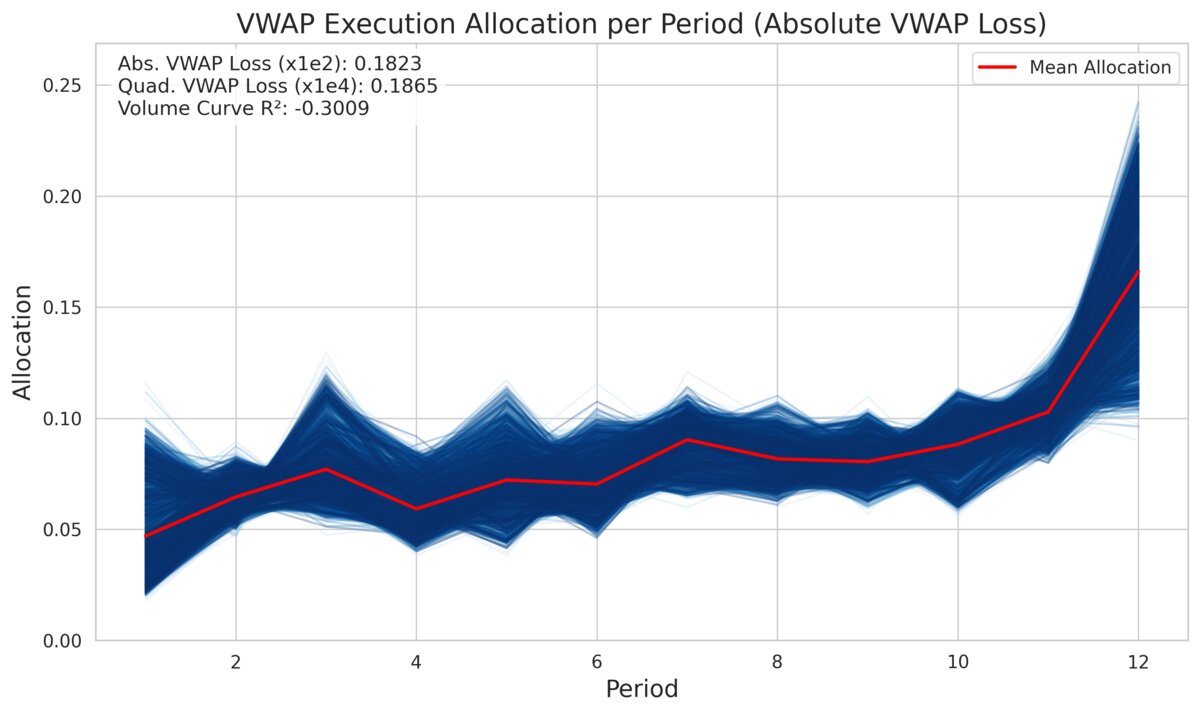}
    \caption{Prediction obtained with model calibrated using Absolute Deviation Loss}
    \label{fig:static_prediction_absolute}
\end{figure}

\begin{figure}[H]
    \centering
    \includegraphics[width=\columnwidth]{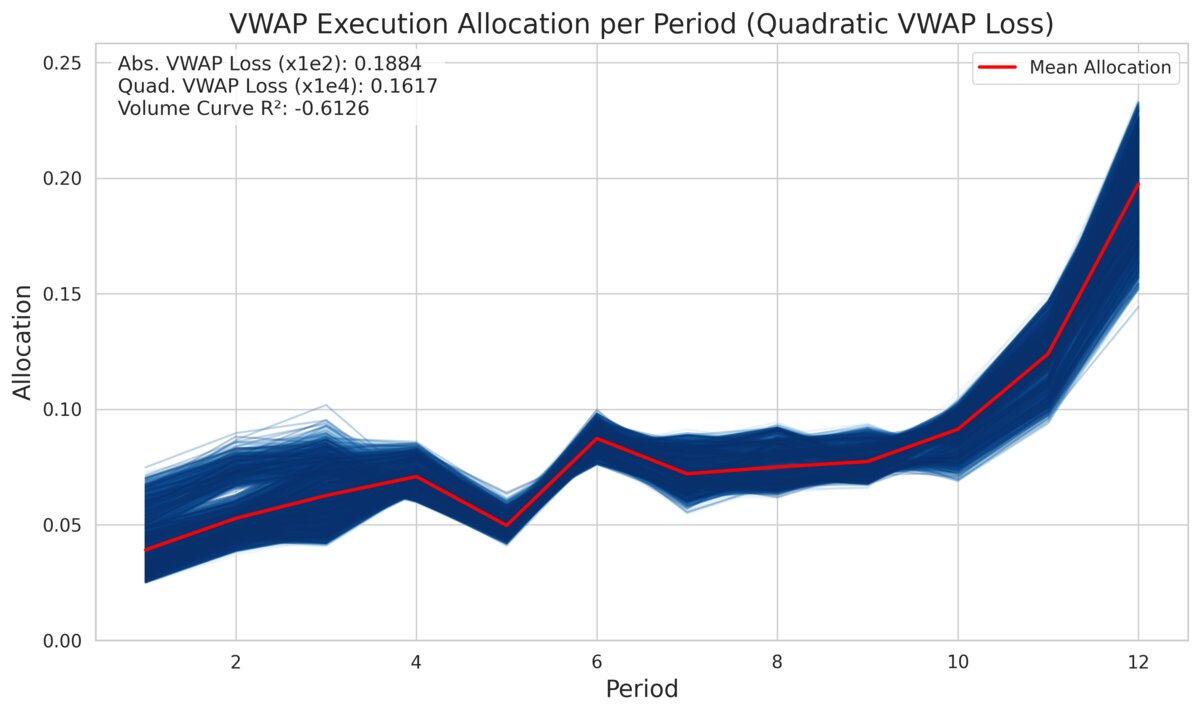}
    \caption{Prediction obtained with model calibrated using Quadratic Deviation Loss}
    \label{fig:static_prediction_quadratic}
\end{figure}

An analysis of the predicted allocation curves under different calibration losses is presented to elucidate how the proposed model differentiates itself from competing approaches. Figures~\ref{fig:static_prediction_absolute} and~\ref{fig:static_prediction_quadratic} illustrate that the predictions generated using absolute and quadratic VWAP losses tend to follow an average allocation pattern similar to that observed in Section~2, while still adapting to current market conditions. Notably, the model trained with quadratic loss produces a more compact prediction region, suggesting that it sacrifices flexibility in favor of an end-loaded execution pattern. In contrast, the model trained with volume curve loss results in a nearly flat allocation, with distinct seasonal variations clearly visible. These visual differences provide important insight into how the choice of loss function shapes the resulting allocation strategy, ultimately contributing to the performance differences observed in the quantitative analysis.

\begin{figure}[H]
    \centering
    \includegraphics[width=\columnwidth]{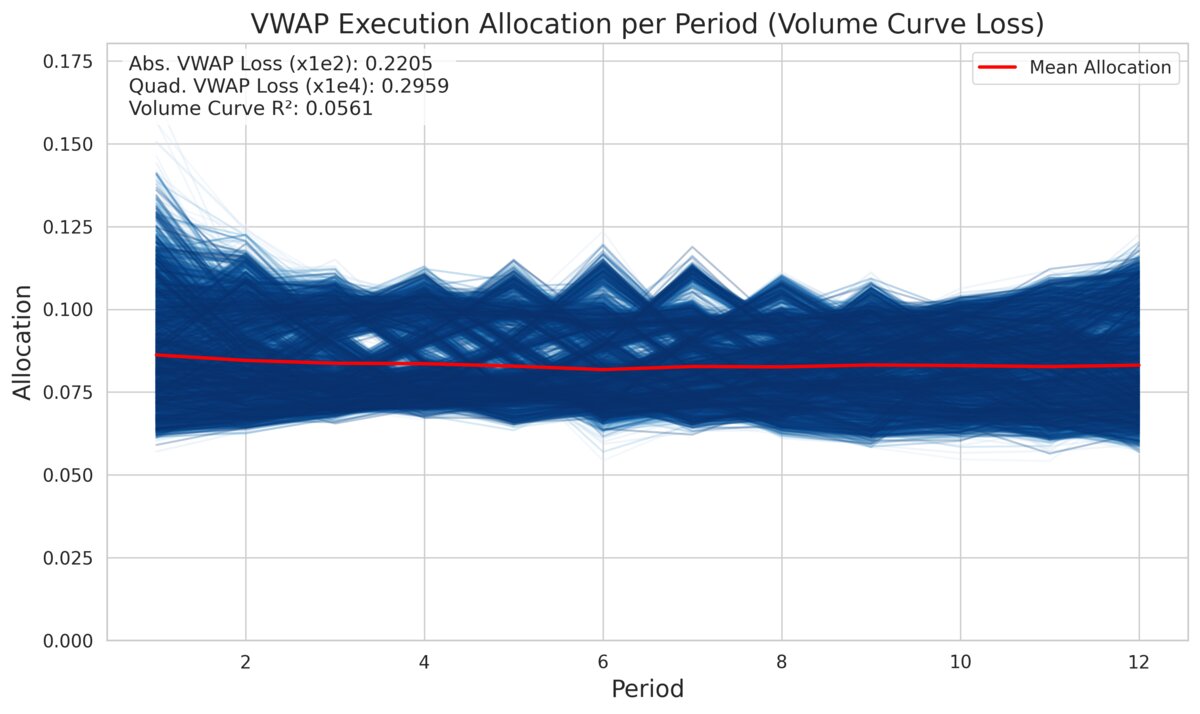}
    \caption{Prediction obtained with model calibrated using Volume Curve Loss}
    \label{fig:static_prediction_volume}
\end{figure}

In contrast, Figure~\ref{fig:static_prediction_volume} shows the allocation curve from a model trained with volume curve loss, which is noticeably flatter and exhibits distinct seasonal waves. These differences underscore that directly optimizing the VWAP objective (via absolute or quadratic loss) yields a strategy that deviates from simply mimicking the historical volume curve.

\subsection{Understanding When It Makes a Difference}
\begin{figure}[H]
    \centering
    \includegraphics[width=\columnwidth]{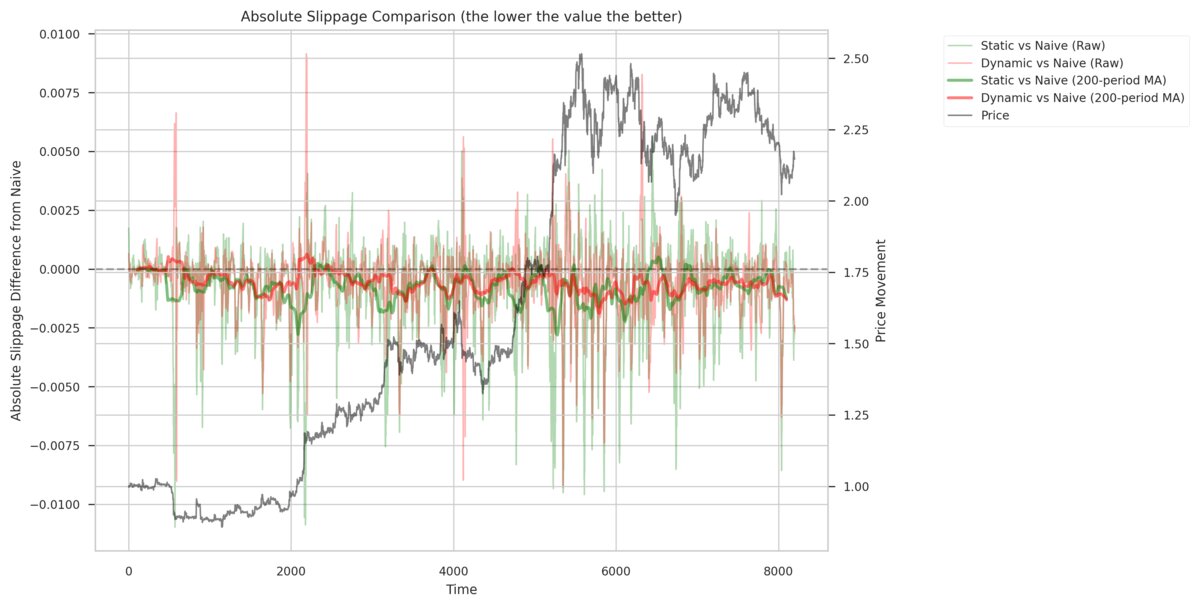}
    \caption{Slippage between approaches on the full out-of-sample set}
    \label{fig:static_slippage_full}
\end{figure}

\begin{figure}[H]
    \centering
    \includegraphics[width=\columnwidth]{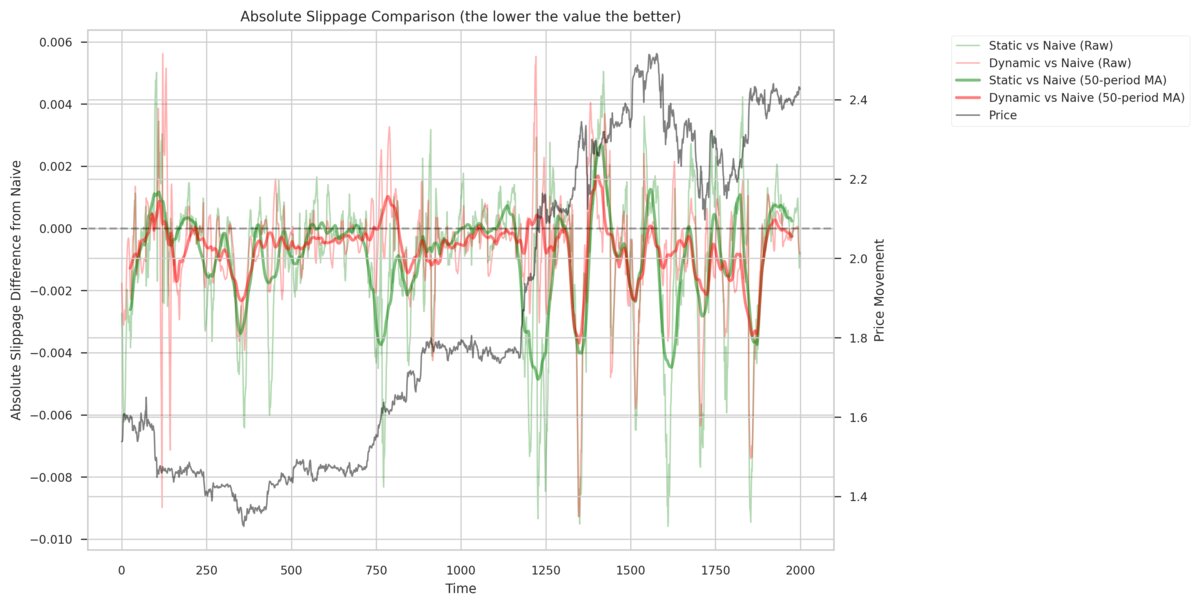}
    \caption{Slippage between approaches on a subsample of the out-of-sample set}
    \label{fig:static_slippage_subset}
\end{figure}

\begin{figure}[H]
    \centering
    \includegraphics[width=\columnwidth]{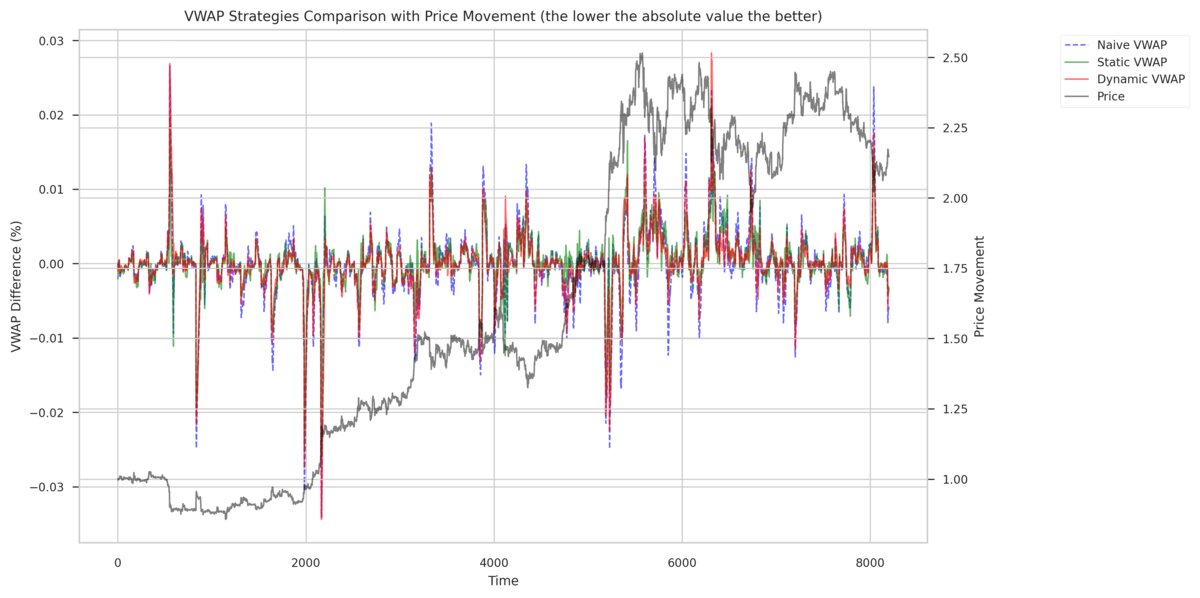}
    \caption{Difference in absolute slippage versus naive approach on the full out-of-sample set. Negative values indicate improved performance over the naive approach.}
    \label{fig:static_slippage_diff_full}
\end{figure}

\begin{figure}[H]
    \centering
    \includegraphics[width=\columnwidth]{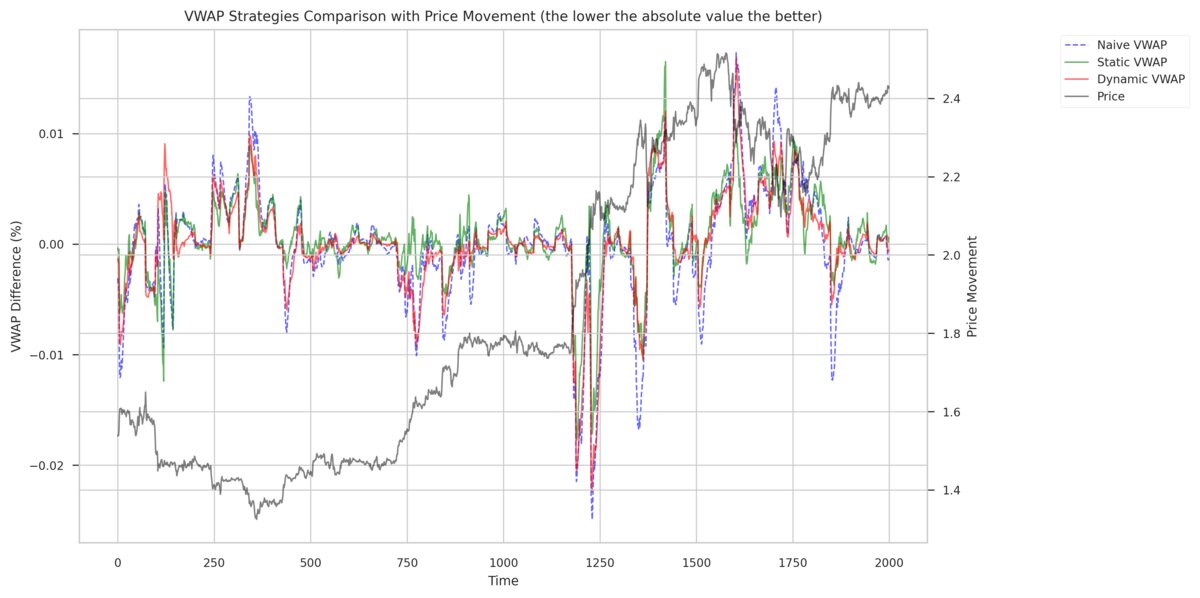}
    \caption{Difference in absolute slippage versus naive approach on the subsample set. Negative values indicate improved performance over the naive approach.}
    \label{fig:static_slippage_diff_subset}
\end{figure}

To further assess the effectiveness of the approach, an analysis of the temporal evolution of slippage (i.e., the difference between the achieved VWAP and the market VWAP) relative to the price is provided. Figure~\ref{fig:static_slippage_full} shows the slippage across the full out-of-sample period for the naive approach, a dynamic VWAP strategy, and the proposed static approach when executing a 48-hour VWAP in hourly bins. Notably, the proposed method generally exhibits lower slippage during periods of high volatility, and in some instances, the slippage even reverses sign relative to standard approaches. To illustrate this effect in greater detail, Figure~\ref{fig:static_slippage_subset} provides a zoomed-in view over a 2000-hour subperiod. The reduced magnitude and frequency of extreme slippage events in the proposed approach are evident in this more focused view. Finally, Figures~\ref{fig:static_slippage_diff_full} and~\ref{fig:static_slippage_diff_subset} present the differences in absolute slippage between the proposed method (or the dynamic linear method) and the naive approach. In these plots, negative values indicate lower absolute slippage (and thus improved performance) relative to the naive strategy, while positive values indicate higher slippage. The prevalence of negative values during volatile periods provides strong visual evidence that the end-loaded allocation strategy is effective in mitigating execution slippage.

\section{Conclusion}

In conclusion, the results of this study demonstrate that directly optimizing the VWAP execution objective through a deep learning framework yields substantial performance improvements over traditional volume curve-based approaches. More broadly, this example illustrates that in many cases where multi-step methods or proxy variables have traditionally been employed, deep learning offers a powerful alternative by enabling direct optimization of the true objective. The expressivity of deep learning allows for straightforward modifications to incorporate any function of market impact, thereby providing a versatile platform for capturing complex trading dynamics.

\medskip

Although a simple, naive linear model (the Temporal Linear Network) was employed, its success is attributable to the fundamentally different approach of targeting the VWAP execution objective directly. This shift in methodology not only enhances execution performance but also underscores the potential of deep learning to transform similar problems in other domains. I believe that this research represents a significant advancement in the design of VWAP execution strategies, and I anticipate that further enhancements—such as dynamic updating mechanisms and the integration of sophisticated market impact models—will extend the applicability of this approach to a broader range of asset classes and trading environments.

\section*{Code Availability}
The source code used for all experiments and analyses in this paper is available at \url{https://github.com/remigenet/DeepLearningVWAP}.

\bibliographystyle{abbrv}
\bibliography{bib}

\addtocontents{toc}{\setcounter{tocdepth}{1}}
\newpage
\appendix
\input{appendix/main}

\end{document}

%% file: tables/2_fix_curve_table.tex
\begin{table}[H]
    \centering
    \caption{VWAP Optimization Results}
    \label{tab:static_fix_curve_vwap_results}
    \small
    \resizebox{\textwidth}{!}{%
        \begin{tabular}{llllcccc}
        \toprule
        \textbf{Asset} & \textbf{Optimization} & \textbf{Method} & \textbf{Abs VWAP Loss} & \textbf{Quad VWAP Loss} & \textbf{Vol Curve Loss} & \textbf{R2 Vol Curve} & \textbf{Opt Time (s)} \\
         & \textbf{Function} &  & \textbf{(1e2)} & \textbf{(1e4)} & \textbf{(1e2)} &  & \\
        \midrule
        \multirow{9}{*}{ADA} & \multirow{3}{*}{absolute\_vwap\_loss} & SLSQP & 0.161262 & 0.103157 & 0.509109 & -0.347896 & 34.626976 \\
         &  & basinhopping & 0.161266 & 0.099737 & 0.556507 & -0.473385 & 55.531589 \\
         &  & DE & 0.159714 & 0.103534 & 0.463983 & -0.228422 & 2.072001 \\
        \cline{2-8}
         & \multirow{3}{*}{quadratic\_vwap\_loss} & SLSQP & 0.161468 & 0.097612 & 0.615348 & -0.629170 & 11.889972 \\
         &  & basinhopping & 0.187822 & 0.102028 & 0.963292 & -1.550373 & 27.556498 \\
         &  & DE & 0.164090 & 0.092467 & 0.576554 & -0.526461 & 3.021690 \\
        \cline{2-8}
         & \multirow{3}{*}{volume\_curve\_loss} & SLSQP & 0.182052 & 0.153098 & 0.379948 & -0.005934 & 25.458261 \\
         &  & basinhopping & 0.184498 & 0.156269 & 0.382820 & -0.013539 & 36.394290 \\
         &  & DE & 0.182067 & 0.152669 & 0.377981 & -0.000728 & 1.807036 \\
        \midrule
        \multirow{9}{*}{BTC} & \multirow{3}{*}{absolute\_vwap\_loss} & SLSQP & 0.105354 & 0.037911 & 0.855709 & -0.628073 & 34.237416 \\
         &  & basinhopping & 0.104696 & 0.037538 & 0.664384 & -0.264057 & 48.594292 \\
         &  & DE & 0.105952 & 0.040566 & 0.636615 & -0.211224 & 2.709536 \\
        \cline{2-8}
         & \multirow{3}{*}{quadratic\_vwap\_loss} & SLSQP & 0.105112 & 0.037638 & 0.728556 & -0.386152 & 12.000814 \\
         &  & basinhopping & 0.108195 & 0.037482 & 1.308381 & -1.489326 & 26.424249 \\
         &  & DE & 0.103572 & 0.035020 & 0.780900 & -0.485740 & 4.301013 \\
        \cline{2-8}
         & \multirow{3}{*}{volume\_curve\_loss} & SLSQP & 0.126713 & 0.061470 & 0.528446 & -0.005422 & 21.426937 \\
         &  & basinhopping & 0.122180 & 0.057882 & 0.535264 & -0.018393 & 34.317264 \\
         &  & DE & 0.124038 & 0.058830 & 0.525396 & 0.000381 & 1.163200 \\
        \midrule
        \multirow{9}{*}{ETH} & \multirow{3}{*}{absolute\_vwap\_loss} & SLSQP & 0.121005 & 0.053397 & 0.696746 & -0.418203 & 36.342283 \\
         &  & basinhopping & 0.125103 & 0.059826 & 0.680083 & -0.384285 & 52.675163 \\
         &  & DE & 0.120213 & 0.053212 & 0.622947 & -0.267986 & 2.533839 \\
        \cline{2-8}
         & \multirow{3}{*}{quadratic\_vwap\_loss} & SLSQP & 0.120978 & 0.052524 & 0.709165 & -0.443481 & 11.941791 \\
         &  & basinhopping & 0.128118 & 0.052153 & 0.881157 & -0.793564 & 27.980689 \\
         &  & DE & 0.119349 & 0.049538 & 0.678544 & -0.381152 & 3.884477 \\
        \cline{2-8}
         & \multirow{3}{*}{volume\_curve\_loss} & SLSQP & 0.147338 & 0.084255 & 0.493581 & -0.004668 & 25.072747 \\
         &  & basinhopping & 0.143035 & 0.080341 & 0.491345 & -0.000117 & 33.726844 \\
         &  & DE & 0.144048 & 0.081395 & 0.491395 & -0.000218 & 2.017230 \\
        \bottomrule
        \end{tabular}
    }
    \caption*{Note: Abs VWAP Loss and Vol Curve Loss are presented in $10^2$ scale, Quad VWAP Loss is presented in $10^4$ scale as per the original data.}
\end{table}

%% file: tables/5_results_table.tex
\begin{table}[H]
    \centering
    \caption{VWAP Optimization Results for 12 steps ahead and 120 lookback window}
    \small
    \resizebox{\textwidth}{!}{%
        \begin{tabular}{llcccccccccc}
        \hline
        Model Type & Asset & Optimization & \multicolumn{2}{c}{Abs. VWAP Loss ($10^{-2}$)} & \multicolumn{2}{c}{Quad. VWAP Loss ($10^{-4}$)} & \multicolumn{2}{c}{R² Vol. Curve} & \multicolumn{2}{c}{Training Time (s)} \\
         &  & Function & Mean & Std & Mean & Std & Mean & Std & Mean & Std \\
        \hline
        Naive & BTC & N/A & 0.158743 & 0.000000 & 0.087808 & 0.000000 & 0.000000 & 0.000000 & 0.000000 & 0.000000 \\
        StaticVWAP & BTC & Absolute & \textbf{0.119742} & 0.000835 & 0.050175 & 0.001045 & -0.133930 & 0.038680 & 17.066662 & 1.452658 \\
        StaticVWAP & BTC & Quadratic & 0.120810 & 0.000759 & \textbf{0.047403} & 0.001055 & -0.361577 & 0.082153 & 16.568835 & 0.591146 \\
        StaticVWAP & BTC & Volume & 0.149938 & 0.001896 & 0.084989 & 0.001637 & 0.134904 & 0.011795 & 24.140719 & 3.765817 \\
        Dynamic Linear & BTC & N/A & 0.142535 & 0.000000 & 0.085731 & 0.000000 & \textbf{0.193757} & 0.000000 & 0.241681 & 0.000000 \\
        Static Linear & BTC & N/A & 0.146979 & 0.000000 & 0.084170 & 0.000000 & 0.162770 & 0.000000 & 0.244588 & 0.000000 \\
        Dynamic StaticVWAP & BTC & Absolute & 0.142334 & 0.008724 & 0.076042 & 0.010107 & 0.021618 & 0.038493 & 15.005514 & 0.799850 \\
        Dynamic StaticVWAP & BTC & Quadratic & 0.147608 & 0.018183 & 0.074102 & 0.016007 & -0.495363 & 0.815728 & 14.517617 & 0.582770 \\
        Dynamic StaticVWAP & BTC & Volume & 0.152135 & 0.004588 & 0.091675 & 0.003997 & 0.148551 & 0.018475 & 20.605638 & 2.945344 \\
        Fixed Volume Curve & BTC & Absolute & 0.129670 & 0.000000 & 0.054213 & 0.000000 & -0.273680 & 0.000000 & 17.263421 & 0.000000 \\
        Fixed Volume Curve & BTC & Quadratic & 0.127530 & 0.000000 & 0.049655 & 0.000000 & -0.467000 & 0.000000 & 30.385431 & 0.000000 \\
        Fixed Volume Curve & BTC & Volume & 0.160575 & 0.000000 & 0.089770 & 0.000000 & -0.003349 & 0.000000 & 12.861315 & 0.000000 \\
        \hline
        Naive & ETH & N/A & 0.177758 & 0.000000 & 0.116196 & 0.000000 & 0.000000 & 0.000000 & 0.000000 & 0.000000 \\
        StaticVWAP & ETH & Absolute & \textbf{0.138627} & 0.000675 & 0.076506 & 0.001128 & -0.146691 & 0.026960 & 16.977891 & 1.268639 \\
        StaticVWAP & ETH & Quadratic & 0.139999 & 0.000451 & 0.073385 & 0.001008 & -0.297154 & 0.051581 & 16.011362 & 0.614544 \\
        StaticVWAP & ETH & Volume & 0.170102 & 0.001785 & 0.122055 & 0.001819 & 0.109135 & 0.007899 & 19.888228 & 3.165192 \\
        Dynamic Linear & ETH & N/A & 0.158829 & 0.000000 & 0.121279 & 0.000000 & \textbf{0.169994} & 0.000000 & 0.249899 & 0.000000 \\
        Static Linear & ETH & N/A & 0.166326 & 0.000000 & 0.122063 & 0.000000 & 0.135824 & 0.000000 & 0.278128 & 0.000000 \\
        Dynamic StaticVWAP & ETH & Absolute & 0.152641 & 0.003763 & 0.093910 & 0.005712 & -0.119677 & 0.128250 & 15.180526 & 1.168761 \\
        Dynamic StaticVWAP & ETH & Quadratic & 0.162467 & 0.017372 & 0.094459 & 0.013988 & -0.633158 & 0.571764 & 14.129841 & 0.686132 \\
        Dynamic StaticVWAP & ETH & Volume & 0.171854 & 0.004362 & 0.127756 & 0.004184 & 0.118685 & 0.018637 & 17.855957 & 2.545837 \\
        Fixed Volume Curve & ETH & Absolute & 0.148407 & 0.000000 & 0.076206 & 0.000000 & -0.269217 & 0.000000 & 16.648768 & 0.000000 \\
        Fixed Volume Curve & ETH & Quadratic & 0.147058 & 0.000000 & \textbf{0.072852} & 0.000000 & -0.348835 & 0.000000 & 23.099389 & 0.000000 \\
        Fixed Volume Curve & ETH & Volume & 0.174744 & 0.000000 & 0.112633 & 0.000000 & -0.006033 & 0.000000 & 9.376649 & 0.000000 \\
        \hline
        Naive & BNB & N/A & 0.174116 & 0.000000 & 0.118874 & 0.000000 & 0.000000 & 0.000000 & 0.000000 & 0.000000 \\
        StaticVWAP & BNB & Absolute & \textbf{0.137741} & 0.000559 & 0.076553 & 0.001233 & -0.169736 & 0.036491 & 15.882844 & 0.794736 \\
        StaticVWAP & BNB & Quadratic & 0.141483 & 0.000963 & \textbf{0.071090} & 0.000803 & -0.477507 & 0.071598 & 15.598018 & 0.652385 \\
        StaticVWAP & BNB & Volume & 0.165025 & 0.001455 & 0.116136 & 0.001633 & 0.086762 & 0.002815 & 18.406203 & 1.898933 \\
        Dynamic Linear & BNB & N/A & 0.160003 & 0.000000 & 0.115190 & 0.000000 & \textbf{0.127731} & 0.000000 & 0.262799 & 0.000000 \\
        Static Linear & BNB & N/A & 0.163049 & 0.000000 & 0.114673 & 0.000000 & 0.101505 & 0.000000 & 0.216979 & 0.000000 \\
        Dynamic StaticVWAP & BNB & Absolute & 0.146537 & 0.006959 & 0.074951 & 0.002269 & -0.434465 & 0.221577 & 14.089177 & 0.626975 \\
        Dynamic StaticVWAP & BNB & Quadratic & 0.183354 & 0.040719 & 0.098591 & 0.031602 & -1.490792 & 1.373252 & 13.962810 & 0.606966 \\
        Dynamic StaticVWAP & BNB & Volume & 0.165976 & 0.003899 & 0.119578 & 0.003757 & 0.093674 & 0.009887 & 16.622247 & 1.866124 \\
        Fixed Volume Curve & BNB & Absolute & 0.146639 & 0.000000 & 0.079064 & 0.000000 & -0.283410 & 0.000000 & 16.226866 & 0.000000 \\
        Fixed Volume Curve & BNB & Quadratic & 0.147920 & 0.000000 & 0.074117 & 0.000000 & -0.440821 & 0.000000 & 31.472387 & 0.000000 \\
        Fixed Volume Curve & BNB & Volume & 0.177473 & 0.000000 & 0.122219 & 0.000000 & -0.003080 & 0.000000 & 12.545491 & 0.000000 \\
        \hline
        Naive & ADA & N/A & 0.226207 & 0.000000 & 0.228809 & 0.000000 & 0.000000 & 0.000000 & 0.000000 & 0.000000 \\
        StaticVWAP & ADA & Absolute & \textbf{0.187574} & 0.000640 & 0.156983 & 0.004485 & -0.201991 & 0.040232 & 15.773802 & 0.727374 \\
        StaticVWAP & ADA & Quadratic & 0.194378 & 0.002171 & 0.139060 & 0.002399 & -0.503548 & 0.086766 & 15.721992 & 0.543023 \\
        StaticVWAP & ADA & Volume & 0.219882 & 0.001456 & 0.244318 & 0.004173 & 0.089412 & 0.004083 & 18.606664 & 2.294395 \\
        Dynamic Linear & ADA & N/A & 0.213691 & 0.000000 & 0.258357 & 0.000000 & \textbf{0.142651} & 0.000000 & 0.236307 & 0.000000 \\
        Static Linear & ADA & N/A & 0.217345 & 0.000000 & 0.253104 & 0.000000 & 0.112113 & 0.000000 & 0.293910 & 0.000000 \\
        Dynamic StaticVWAP & ADA & Absolute & 0.224430 & 0.018242 & 0.164912 & 0.013876 & -0.868068 & 0.389384 & 14.117266 & 0.803708 \\
        Dynamic StaticVWAP & ADA & Quadratic & 0.235337 & 0.043516 & 0.168912 & 0.041709 & -1.121577 & 1.028339 & 13.869841 & 0.524214 \\
        Dynamic StaticVWAP & ADA & Volume & 0.223166 & 0.005122 & 0.253597 & 0.009979 & 0.094306 & 0.010872 & 16.104520 & 2.195146 \\
        Fixed Volume Curve & ADA & Absolute & 0.195395 & 0.000000 & 0.148712 & 0.000000 & -0.311109 & 0.000000 & 15.779793 & 0.000000 \\
        Fixed Volume Curve & ADA & Quadratic & 0.199729 & 0.000000 & \textbf{0.135216} & 0.000000 & -0.573288 & 0.000000 & 22.675063 & 0.000000 \\
        Fixed Volume Curve & ADA & Volume & 0.231114 & 0.000000 & 0.236772 & 0.000000 & -0.005151 & 0.000000 & 11.179776 & 0.000000 \\
        \hline
        Naive & XRP & N/A & 0.223855 & 0.000000 & 0.275925 & 0.000000 & 0.000000 & 0.000000 & 0.000000 & 0.000000 \\
        StaticVWAP & XRP & Absolute & \textbf{0.182436} & 0.000558 & 0.187474 & 0.003841 & -0.268477 & 0.054168 & 16.156652 & 0.781386 \\
        StaticVWAP & XRP & Quadratic & 0.188880 & 0.001564 & 0.158377 & 0.003440 & -0.800538 & 0.151472 & 15.717826 & 0.651796 \\
        StaticVWAP & XRP & Volume & 0.218816 & 0.001646 & 0.291965 & 0.003142 & 0.053552 & 0.002636 & 18.000321 & 2.353914 \\
        Dynamic Linear & XRP & N/A & 0.213728 & 0.000000 & 0.295177 & 0.000000 & \textbf{0.098195} & 0.000000 & 0.290283 & 0.000000 \\
        Static Linear & XRP & N/A & 0.216581 & 0.000000 & 0.295605 & 0.000000 & 0.071538 & 0.000000 & 0.248654 & 0.000000 \\
        Dynamic StaticVWAP & XRP & Absolute & 0.213859 & 0.017346 & 0.169141 & 0.009132 & -1.273806 & 0.583631 & 14.198917 & 0.578075 \\
        Dynamic StaticVWAP & XRP & Quadratic & 0.249005 & 0.038793 & 0.195797 & 0.028412 & -2.392281 & 1.708965 & 13.999556 & 0.594463 \\
        Dynamic StaticVWAP & XRP & Volume & 0.221568 & 0.003417 & 0.299404 & 0.007929 & 0.052095 & 0.005762 & 15.439939 & 1.495954 \\
        Fixed Volume Curve & XRP & Absolute & 0.186341 & 0.000000 & 0.178947 & 0.000000 & -0.347910 & 0.000000 & 14.544090 & 0.000000 \\
        Fixed Volume Curve & XRP & Quadratic & 0.189142 & 0.000000 & \textbf{0.152440} & 0.000000 & -0.794792 & 0.000000 & 23.787354 & 0.000000 \\
        Fixed Volume Curve & XRP & Volume & 0.224941 & 0.000000 & 0.276570 & 0.000000 & -0.001512 & 0.000000 & 12.143459 & 0.000000 \\
        \hline
        \end{tabular}
    }
    \label{tab:static_vwap_results}
\end{table}

%% file: appendix/main.tex
\section{Results Tables}
\input{appendix/tables/main}

\input{appendix/figures/main}

%% file: appendix/tables/main.tex
\input{appendix/tables/result_6_ahead}
\input{appendix/tables/results_48_ahead}

%% file: appendix/tables/result_6_ahead.tex
\begin{table}[H]
    \centering
    \caption{VWAP Optimization Results for 6 steps ahead and 15 lookback window}
    \small
    \resizebox{\textwidth}{!}{%
        \begin{tabular}{llcccccccccc}
        \hline
        Model Type & Asset & Optimization & \multicolumn{2}{c}{Abs. VWAP Loss ($10^{-2}$)} & \multicolumn{2}{c}{Quad. VWAP Loss ($10^{-4}$)} & \multicolumn{2}{c}{R² Vol. Curve} & \multicolumn{2}{c}{Training Time (s)} \\
         &  & Function & Mean & Std & Mean & Std & Mean & Std & Mean & Std \\
        \hline
        Naive & BTC & N/A & 0.105363 & 0.000000 & 0.046562 & 0.000000 & 0.000000 & 0.000000 & 0.000000 & 0.000000 \\
        StaticVWAP & BTC & Absolute & \textbf{0.087892} & 0.000429 & 0.030610 & 0.000547 & -0.183722 & 0.031797 & 12.667258 & 0.715114 \\
        StaticVWAP & BTC & Quadratic & 0.089215 & 0.000280 & \textbf{0.027780} & 0.000445 & -0.534662 & 0.054181 & 12.640691 & 0.591074 \\
        StaticVWAP & BTC & Volume & 0.103216 & 0.000442 & 0.048582 & 0.000407 & 0.081896 & 0.001127 & 17.762622 & 2.642843 \\
        Dynamic Linear & BTC & N/A & 0.100858 & 0.000000 & 0.048049 & 0.000000 & \textbf{0.109054} & 0.000000 & 0.049620 & 0.000000 \\
        Static Linear & BTC & N/A & 0.102500 & 0.000000 & 0.048115 & 0.000000 & 0.087037 & 0.000000 & 0.035565 & 0.000000 \\
        Dynamic StaticVWAP & BTC & Absolute & 0.094378 & 0.001417 & 0.038101 & 0.001182 & -0.024425 & 0.016048 & 11.196409 & 0.682693 \\
        Dynamic StaticVWAP & BTC & Quadratic & 0.098605 & 0.004904 & 0.041420 & 0.004580 & -0.060975 & 0.034466 & 11.247421 & 0.582951 \\
        Dynamic StaticVWAP & BTC & Volume & 0.102808 & 0.000661 & 0.049507 & 0.000656 & 0.095136 & 0.001851 & 14.932640 & 2.098674 \\
        Fixed Volume Curve & BTC & Absolute & 0.090429 & 0.000000 & 0.030996 & 0.000000 & -0.233945 & 0.000000 & 5.167805 & 0.000000 \\
        Fixed Volume Curve & BTC & Quadratic & 0.090701 & 0.000000 & 0.028026 & 0.000000 & -0.543176 & 0.000000 & 6.107834 & 0.000000 \\
        Fixed Volume Curve & BTC & Volume & 0.105142 & 0.000000 & 0.046175 & 0.000000 & 0.000909 & 0.000000 & 4.947306 & 0.000000 \\
        \hline
        Naive & ETH & N/A & 0.121152 & 0.000000 & 0.061690 & 0.000000 & 0.000000 & 0.000000 & 0.000000 & 0.000000 \\
        StaticVWAP & ETH & Absolute & \textbf{0.100699} & 0.000554 & 0.042535 & 0.000771 & -0.171298 & 0.026887 & 12.739415 & 1.228726 \\
        StaticVWAP & ETH & Quadratic & 0.101188 & 0.000298 & 0.039114 & 0.000351 & -0.438104 & 0.043960 & 12.285954 & 0.616164 \\
        StaticVWAP & ETH & Volume & 0.117997 & 0.000835 & 0.065511 & 0.000817 & 0.073627 & 0.001178 & 16.056946 & 2.581432 \\
        Dynamic Linear & ETH & N/A & 0.114201 & 0.000000 & 0.063881 & 0.000000 & \textbf{0.103432} & 0.000000 & 0.034144 & 0.000000 \\
        Static Linear & ETH & N/A & 0.117400 & 0.000000 & 0.065136 & 0.000000 & 0.075838 & 0.000000 & 0.024283 & 0.000000 \\
        Dynamic StaticVWAP & ETH & Absolute & 0.107189 & 0.001218 & 0.050199 & 0.001201 & -0.024604 & 0.016778 & 10.827399 & 0.818096 \\
        Dynamic StaticVWAP & ETH & Quadratic & 0.108867 & 0.005334 & 0.050609 & 0.005237 & -0.090708 & 0.049720 & 10.975629 & 0.566742 \\
        Dynamic StaticVWAP & ETH & Volume & 0.116336 & 0.001107 & 0.065519 & 0.001174 & 0.091956 & 0.002112 & 14.591255 & 2.019878 \\
        Fixed Volume Curve & ETH & Absolute & 0.103074 & 0.000000 & 0.041647 & 0.000000 & -0.235016 & 0.000000 & 5.236793 & 0.000000 \\
        Fixed Volume Curve & ETH & Quadratic & 0.102993 & 0.000000 & \textbf{0.038466} & 0.000000 & -0.501563 & 0.000000 & 7.602352 & 0.000000 \\
        Fixed Volume Curve & ETH & Volume & 0.121202 & 0.000000 & 0.061560 & 0.000000 & 0.000962 & 0.000000 & 10.572813 & 0.000000 \\
        \hline
        Naive & BNB & N/A & 0.114749 & 0.000000 & 0.058204 & 0.000000 & 0.000000 & 0.000000 & 0.000000 & 0.000000 \\
        StaticVWAP & BNB & Absolute & \textbf{0.098716} & 0.000371 & 0.040009 & 0.000516 & -0.179600 & 0.021833 & 11.748436 & 0.604826 \\
        StaticVWAP & BNB & Quadratic & 0.102098 & 0.000456 & \textbf{0.037119} & 0.000263 & -0.555685 & 0.036628 & 11.970830 & 0.563751 \\
        StaticVWAP & BNB & Volume & 0.111265 & 0.000567 & 0.059603 & 0.000614 & 0.057254 & 0.000958 & 17.131891 & 2.764334 \\
        Dynamic Linear & BNB & N/A & 0.109450 & 0.000000 & 0.060628 & 0.000000 & \textbf{0.083574} & 0.000000 & 0.021749 & 0.000000 \\
        Static Linear & BNB & N/A & 0.111120 & 0.000000 & 0.060123 & 0.000000 & 0.061241 & 0.000000 & 0.021126 & 0.000000 \\
        Dynamic StaticVWAP & BNB & Absolute & 0.100140 & 0.000957 & 0.043232 & 0.001259 & -0.076393 & 0.026245 & 10.566655 & 0.743802 \\
        Dynamic StaticVWAP & BNB & Quadratic & 0.102059 & 0.003182 & 0.043621 & 0.004242 & -0.168157 & 0.100905 & 10.780092 & 0.561580 \\
        Dynamic StaticVWAP & BNB & Volume & 0.110698 & 0.001120 & 0.061614 & 0.001189 & 0.069469 & 0.007095 & 13.911978 & 1.804252 \\
        Fixed Volume Curve & BNB & Absolute & 0.100777 & 0.000000 & 0.040704 & 0.000000 & -0.229291 & 0.000000 & 5.064288 & 0.000000 \\
        Fixed Volume Curve & BNB & Quadratic & 0.102581 & 0.000000 & 0.037567 & 0.000000 & -0.554224 & 0.000000 & 7.126949 & 0.000000 \\
        Fixed Volume Curve & BNB & Volume & 0.113358 & 0.000000 & 0.056691 & 0.000000 & 0.000029 & 0.000000 & 3.750299 & 0.000000 \\
        \hline
        Naive & ADA & N/A & 0.152054 & 0.000000 & 0.113711 & 0.000000 & 0.000000 & 0.000000 & 0.000000 & 0.000000 \\
        StaticVWAP & ADA & Absolute & \textbf{0.132755} & 0.000331 & 0.081472 & 0.001282 & -0.193195 & 0.020878 & 11.714828 & 0.480636 \\
        StaticVWAP & ADA & Quadratic & 0.138740 & 0.000839 & 0.073484 & 0.000743 & -0.573241 & 0.046742 & 12.054475 & 0.541245 \\
        StaticVWAP & ADA & Volume & 0.148776 & 0.000583 & 0.122563 & 0.001312 & 0.058059 & 0.002759 & 16.160650 & 2.211096 \\
        Dynamic Linear & ADA & N/A & 0.145723 & 0.000000 & 0.123286 & 0.000000 & \textbf{0.089187} & 0.000000 & 0.018582 & 0.000000 \\
        Static Linear & ADA & N/A & 0.147620 & 0.000000 & 0.120900 & 0.000000 & 0.065587 & 0.000000 & 0.018901 & 0.000000 \\
        Dynamic StaticVWAP & ADA & Absolute & 0.133511 & 0.000581 & 0.082349 & 0.002144 & -0.156369 & 0.034254 & 10.611309 & 0.607574 \\
        Dynamic StaticVWAP & ADA & Quadratic & 0.137696 & 0.004007 & 0.089385 & 0.007844 & -0.134158 & 0.055278 & 10.694741 & 0.564848 \\
        Dynamic StaticVWAP & ADA & Volume & 0.147723 & 0.001160 & 0.123852 & 0.001922 & 0.072364 & 0.005116 & 13.739881 & 1.852826 \\
        Fixed Volume Curve & ADA & Absolute & 0.134541 & 0.000000 & 0.077918 & 0.000000 & -0.249498 & 0.000000 & 5.076644 & 0.000000 \\
        Fixed Volume Curve & ADA & Quadratic & 0.139507 & 0.000000 & \textbf{0.070773} & 0.000000 & -0.683673 & 0.000000 & 6.158613 & 0.000000 \\
        Fixed Volume Curve & ADA & Volume & 0.151185 & 0.000000 & 0.112242 & 0.000000 & -0.000125 & 0.000000 & 5.369296 & 0.000000 \\
        \hline
        Naive & XRP & N/A & 0.144983 & 0.000000 & 0.145303 & 0.000000 & 0.000000 & 0.000000 & 0.000000 & 0.000000 \\
        StaticVWAP & XRP & Absolute & \textbf{0.125194} & 0.000205 & 0.103814 & 0.001744 & -0.234420 & 0.027725 & 11.967038 & 0.506264 \\
        StaticVWAP & XRP & Quadratic & 0.131908 & 0.000899 & 0.088307 & 0.001263 & -0.838946 & 0.077896 & 12.080394 & 0.594669 \\
        StaticVWAP & XRP & Volume & 0.142387 & 0.000941 & 0.150526 & 0.001568 & 0.042126 & 0.004917 & 13.581942 & 2.835409 \\
        Dynamic Linear & XRP & N/A & 0.140321 & 0.000000 & 0.150192 & 0.000000 & \textbf{0.066619} & 0.000000 & 0.019342 & 0.000000 \\
        Static Linear & XRP & N/A & 0.141379 & 0.000000 & 0.148847 & 0.000000 & 0.047692 & 0.000000 & 0.023341 & 0.000000 \\
        Dynamic StaticVWAP & XRP & Absolute & 0.126336 & 0.000554 & 0.102693 & 0.003199 & -0.261795 & 0.057650 & 10.709266 & 0.465655 \\
        Dynamic StaticVWAP & XRP & Quadratic & 0.133498 & 0.008056 & 0.101611 & 0.014050 & -0.592432 & 0.456198 & 10.783978 & 0.391321 \\
        Dynamic StaticVWAP & XRP & Volume & 0.142820 & 0.001962 & 0.152208 & 0.002334 & 0.047364 & 0.011307 & 11.902569 & 1.252472 \\
        Fixed Volume Curve & XRP & Absolute & 0.126416 & 0.000000 & 0.100266 & 0.000000 & -0.282541 & 0.000000 & 4.858923 & 0.000000 \\
        Fixed Volume Curve & XRP & Quadratic & 0.130682 & 0.000000 & \textbf{0.084545} & 0.000000 & -0.734930 & 0.000000 & 7.530367 & 0.000000 \\
        Fixed Volume Curve & XRP & Volume & 0.143751 & 0.000000 & 0.142603 & 0.000000 & 0.000280 & 0.000000 & 4.148712 & 0.000000 \\
        \hline
        \end{tabular}
    }
    \label{tab:static_vwap_results_6}
\end{table}

%% file: appendix/tables/results_48_ahead.tex
\begin{table}[H]
    \centering
    \caption{VWAP Optimization Results for 48 steps ahead and 120 lookback window}
    \small
    \resizebox{\textwidth}{!}{%
        \begin{tabular}{llcccccccccc}
        \hline
        Model Type & Asset & Optimization & \multicolumn{2}{c}{Abs. VWAP Loss ($10^{-2}$)} & \multicolumn{2}{c}{Quad. VWAP Loss ($10^{-4}$)} & \multicolumn{2}{c}{R² Vol. Curve} & \multicolumn{2}{c}{Training Time (s)} \\
         &  & Function & Mean & Std & Mean & Std & Mean & Std & Mean & Std \\
        \hline
        Naive & BTC & N/A & 0.294555 & 0.000000 & 0.240161 & 0.000000 & 0.000000 & 0.000000 & 0.000000 & 0.000000 \\
        StaticVWAP & BTC & Absolute & \textbf{0.211433} & 0.004159 & 0.120992 & 0.005565 & -0.164033 & 0.183986 & 14.939717 & 1.717610 \\
        StaticVWAP & BTC & Quadratic & 0.215783 & 0.002679 & \textbf{0.119468} & 0.003441 & -0.603772 & 0.287578 & 13.705642 & 0.736571 \\
        StaticVWAP & BTC & Volume & 0.271076 & 0.004552 & 0.211014 & 0.006431 & 0.156645 & 0.007209 & 13.847590 & 0.600646 \\
        Dynamic Linear & BTC & N/A & 0.255156 & 0.000000 & 0.214156 & 0.000000 & \textbf{0.259255} & 0.000000 & 0.266352 & 0.000000 \\
        Static Linear & BTC & N/A & 0.265879 & 0.000000 & 0.206610 & 0.000000 & 0.193961 & 0.000000 & 0.262161 & 0.000000 \\
        Dynamic StaticVWAP & BTC & Absolute & 0.315446 & 0.115326 & 0.241844 & 0.201273 & -1.847108 & 3.194064 & 13.079563 & 1.249241 \\
        Dynamic StaticVWAP & BTC & Quadratic & 0.444085 & 0.181170 & 0.462257 & 0.351248 & -3.587434 & 4.243950 & 12.183085 & 0.628731 \\
        Dynamic StaticVWAP & BTC & Volume & 0.284008 & 0.013071 & 0.234316 & 0.020490 & 0.163476 & 0.017579 & 12.310637 & 0.828832 \\
        Fixed Volume Curve & BTC & Absolute & 0.232173 & 0.000000 & 0.139101 & 0.000000 & -0.320558 & 0.000000 & 203.007813 & 0.000000 \\
        Fixed Volume Curve & BTC & Quadratic & 0.223227 & 0.000000 & 0.119850 & 0.000000 & -0.404871 & 0.000000 & 457.121906 & 0.000000 \\
        Fixed Volume Curve & BTC & Volume & 0.299372 & 0.000000 & 0.245986 & 0.000000 & -0.014183 & 0.000000 & 233.146574 & 0.000000 \\
        \hline
        Naive & ETH & N/A & 0.306788 & 0.000000 & 0.298487 & 0.000000 & 0.000000 & 0.000000 & 0.000000 & 0.000000 \\
        StaticVWAP & ETH & Absolute & \textbf{0.231228} & 0.002613 & 0.175601 & 0.005331 & -0.330961 & 0.214794 & 13.861206 & 0.746753 \\
        StaticVWAP & ETH & Quadratic & 0.241180 & 0.003945 & 0.178887 & 0.007492 & -0.476988 & 0.332822 & 13.535372 & 0.576845 \\
        StaticVWAP & ETH & Volume & 0.290899 & 0.006168 & 0.290979 & 0.009792 & 0.129610 & 0.009962 & 13.552314 & 0.655621 \\
        Dynamic Linear & ETH & N/A & 0.263310 & 0.000000 & 0.289951 & 0.000000 & \textbf{0.229208} & 0.000000 & 0.265109 & 0.000000 \\
        Static Linear & ETH & N/A & 0.284703 & 0.000000 & 0.289589 & 0.000000 & 0.160990 & 0.000000 & 0.340382 & 0.000000 \\
        Dynamic StaticVWAP & ETH & Absolute & 0.499737 & 0.177198 & 0.553371 & 0.373608 & -3.896599 & 3.884830 & 12.327220 & 0.820315 \\
        Dynamic StaticVWAP & ETH & Quadratic & 0.647119 & 0.234991 & 0.928275 & 0.630363 & -6.944855 & 8.297873 & 11.924670 & 0.644082 \\
        Dynamic StaticVWAP & ETH & Volume & 0.292858 & 0.013695 & 0.302941 & 0.022585 & 0.136741 & 0.013402 & 12.005558 & 0.575222 \\
        Fixed Volume Curve & ETH & Absolute & 0.250284 & 0.000000 & 0.183056 & 0.000000 & -0.350797 & 0.000000 & 176.584955 & 0.000000 \\
        Fixed Volume Curve & ETH & Quadratic & 0.246691 & 0.000000 & \textbf{0.167288} & 0.000000 & -0.577755 & 0.000000 & 482.048066 & 0.000000 \\
        Fixed Volume Curve & ETH & Volume & 0.303970 & 0.000000 & 0.295694 & 0.000000 & -0.019532 & 0.000000 & 146.755024 & 0.000000 \\
        \hline
        Naive & BNB & N/A & 0.306506 & 0.000000 & 0.276410 & 0.000000 & 0.000000 & 0.000000 & 0.000000 & 0.000000 \\
        StaticVWAP & BNB & Absolute & \textbf{0.238088} & 0.001673 & \textbf{0.159082} & 0.002518 & -0.283719 & 0.178850 & 14.012146 & 1.414659 \\
        StaticVWAP & BNB & Quadratic & 0.274563 & 0.004827 & 0.194232 & 0.005945 & -0.772361 & 0.345562 & 13.090553 & 0.609440 \\
        StaticVWAP & BNB & Volume & 0.282180 & 0.003849 & 0.245898 & 0.006076 & 0.092449 & 0.006266 & 13.403553 & 0.599346 \\
        Dynamic Linear & BNB & N/A & 0.270936 & 0.000000 & 0.245040 & 0.000000 & \textbf{0.163863} & 0.000000 & 0.244099 & 0.000000 \\
        Static Linear & BNB & N/A & 0.279849 & 0.000000 & 0.247151 & 0.000000 & 0.109303 & 0.000000 & 0.247065 & 0.000000 \\
        Dynamic StaticVWAP & BNB & Absolute & 0.847744 & 0.171242 & 1.565226 & 0.605965 & -13.173198 & 9.265734 & 12.176307 & 1.000202 \\
        Dynamic StaticVWAP & BNB & Quadratic & 0.869293 & 0.184199 & 1.674239 & 0.635109 & -11.551874 & 9.018422 & 11.636941 & 0.632209 \\
        Dynamic StaticVWAP & BNB & Volume & 0.290814 & 0.007511 & 0.263559 & 0.012282 & 0.103465 & 0.012668 & 11.875909 & 0.509390 \\
        Fixed Volume Curve & BNB & Absolute & 0.259528 & 0.000000 & 0.183188 & 0.000000 & -0.417115 & 0.000000 & 192.202802 & 0.000000 \\
        Fixed Volume Curve & BNB & Quadratic & 0.271884 & 0.000000 & 0.188699 & 0.000000 & -0.677903 & 0.000000 & 549.310031 & 0.000000 \\
        Fixed Volume Curve & BNB & Volume & 0.309604 & 0.000000 & 0.280651 & 0.000000 & -0.022514 & 0.000000 & 210.039872 & 0.000000 \\
        \hline
        Naive & ADA & N/A & 0.419024 & 0.000000 & 0.669289 & 0.000000 & 0.000000 & 0.000000 & 0.000000 & 0.000000 \\
        StaticVWAP & ADA & Absolute & \textbf{0.351652} & 0.002676 & 0.441734 & 0.008008 & -0.370866 & 0.213148 & 13.560816 & 0.722914 \\
        StaticVWAP & ADA & Quadratic & 0.368340 & 0.004074 & 0.412750 & 0.009218 & -0.734764 & 0.333143 & 13.196249 & 0.670882 \\
        StaticVWAP & ADA & Volume & 0.408287 & 0.004543 & 0.733465 & 0.015243 & 0.084650 & 0.003713 & 13.204992 & 0.649402 \\
        Dynamic Linear & ADA & N/A & 0.388671 & 0.000000 & 0.792817 & 0.000000 & \textbf{0.172699} & 0.000000 & 0.251263 & 0.000000 \\
        Static Linear & ADA & N/A & 0.405453 & 0.000000 & 0.754683 & 0.000000 & 0.107097 & 0.000000 & 0.249000 & 0.000000 \\
        Dynamic StaticVWAP & ADA & Absolute & 1.271938 & 0.250209 & 2.927095 & 1.148054 & -17.911495 & 11.084118 & 11.941535 & 0.699107 \\
        Dynamic StaticVWAP & ADA & Quadratic & 1.290262 & 0.241712 & 2.988568 & 1.031134 & -14.352830 & 7.704632 & 11.799472 & 0.598161 \\
        Dynamic StaticVWAP & ADA & Volume & 0.421940 & 0.009454 & 0.769364 & 0.033799 & 0.086176 & 0.011173 & 11.787223 & 0.514459 \\
        Fixed Volume Curve & ADA & Absolute & 0.353334 & 0.000000 & 0.401683 & 0.000000 & -0.517208 & 0.000000 & 197.274881 & 0.000000 \\
        Fixed Volume Curve & ADA & Quadratic & 0.373185 & 0.000000 & \textbf{0.381716} & 0.000000 & -0.777649 & 0.000000 & 484.393773 & 0.000000 \\
        Fixed Volume Curve & ADA & Volume & 0.412232 & 0.000000 & 0.646860 & 0.000000 & -0.019907 & 0.000000 & 224.401588 & 0.000000 \\
        \hline
        Naive & XRP & N/A & 0.417425 & 0.000000 & 0.652992 & 0.000000 & 0.000000 & 0.000000 & 0.000000 & 0.000000 \\
        StaticVWAP & XRP & Absolute & \textbf{0.336897} & 0.002525 & 0.394574 & 0.012198 & -0.385838 & 0.207828 & 13.638464 & 0.637926 \\
        StaticVWAP & XRP & Quadratic & 0.366798 & 0.008950 & 0.338236 & 0.015288 & -1.488659 & 0.668829 & 13.407381 & 0.678238 \\
        StaticVWAP & XRP & Volume & 0.405872 & 0.003963 & 0.670118 & 0.009434 & 0.062685 & 0.005969 & 13.312675 & 0.646472 \\
        Dynamic Linear & XRP & N/A & 0.391767 & 0.000000 & 0.675384 & 0.000000 & \textbf{0.135234} & 0.000000 & 0.237720 & 0.000000 \\
        Static Linear & XRP & N/A & 0.400026 & 0.000000 & 0.669650 & 0.000000 & 0.077930 & 0.000000 & 0.268407 & 0.000000 \\
        Dynamic StaticVWAP & XRP & Absolute & 0.829306 & 0.172708 & 1.305078 & 0.532350 & -11.319136 & 7.766351 & 12.344783 & 1.181881 \\
        Dynamic StaticVWAP & XRP & Quadratic & 0.744789 & 0.220680 & 1.118644 & 0.615147 & -8.111042 & 7.154984 & 12.244301 & 1.173284 \\
        Dynamic StaticVWAP & XRP & Volume & 0.416765 & 0.010058 & 0.707899 & 0.030212 & 0.071229 & 0.008034 & 11.917936 & 0.555746 \\
        Fixed Volume Curve & XRP & Absolute & 0.344144 & 0.000000 & 0.380628 & 0.000000 & -0.411449 & 0.000000 & 195.540658 & 0.000000 \\
        Fixed Volume Curve & XRP & Quadratic & 0.357687 & 0.000000 & \textbf{0.328731} & 0.000000 & -0.623001 & 0.000000 & 559.380669 & 0.000000 \\
        Fixed Volume Curve & XRP & Volume & 0.414514 & 0.000000 & 0.642567 & 0.000000 & -0.015077 & 0.000000 & 211.496234 & 0.000000 \\
        \hline
        \end{tabular}
    }
    \label{tab:static_vwap_results_48}
\end{table}

%% file: appendix/figures/main.tex
\onecolumn
\begin{samepage}
    \section{Execution Curves Graphs}
    \label{sec:execution_curves} 
    \vspace{0.5cm}
\end{samepage}

\newcommand{\assetlist}{BTC,ETH,BNB,ADA,XRP}
\newcommand{\stepslist}{6,12,48}

\foreach \steps in \stepslist {
    \foreach \asset in \assetlist {
        \FloatBarrier
        \ifnum\pdfstrcmp{\steps}{6}=0\relax%
        \ifnum\pdfstrcmp{\asset}{BTC}=0\relax%
        \else
            \clearpage
        \fi
        \else
            \clearpage
        \fi
        
        \begin{samepage}
            \subsection{\asset{} Execution Curves with \steps{} Steps Ahead}
            \vspace{0.3cm}
        \end{samepage}
        
        \begin{multicols}{3}
            \begin{figure}[H]
                \centering
                \includegraphics[width=\linewidth]
                {appendix/figures/vwap_execution_allocation_Fixed_Volume_Curve__\asset_\steps_steps.jpg}
                \caption{Fixed Volume Curve (Absolute Loss)}
                \label{fig:fixed_volume_curve_graph_\asset_\steps}
            \end{figure}
    
            \begin{figure}[H]
                \centering
                \includegraphics[width=\linewidth]
                {appendix/figures/vwap_execution_allocation_Fixed_Volume_Curve_Quadratic_Loss__\asset_\steps_steps.jpg}
                \caption{Fixed Volume Curve (Quadratic Loss)}
                \label{fig:fixed_volume_curve_quad_graph_\asset_\steps}
            \end{figure}
    
            \begin{figure}[H]
                \centering
                \includegraphics[width=\linewidth]
                {appendix/figures/vwap_execution_allocation_Fixed_Volume_Curve_Volume_Curve_Loss__\asset_\steps_steps.jpg}
                \caption{Fixed Volume Curve (Volume Curve Loss)}
                \label{fig:fixed_volume_curve_vc_graph_\asset_\steps}
            \end{figure}
            
            \begin{figure}[H]
                \centering
                \includegraphics[width=\linewidth]
                {appendix/figures/vwap_execution_allocation_Static_Linear_Regression__\asset_\steps_steps.jpg}
                \caption{Static Linear Regression}
                \label{fig:linear_regression_graph_\asset_\steps}
            \end{figure}
            
            \begin{figure}[H]
                \centering
                \includegraphics[width=\linewidth]
                {appendix/figures/vwap_execution_allocation_Dynamic_Linear_Regression__\asset_\steps_steps.jpg}
                \caption{Dynamic Linear Regression}
                \label{fig:dynamic_linear_graph_\asset_\steps}
            \end{figure}
            
            \begin{figure}[H]
                \centering
                \includegraphics[width=\linewidth]
                {appendix/figures/vwap_execution_allocation_StaticVWAP_using_absolute_vwap_loss_\asset_\steps_steps.jpg}
                \caption{StaticVWAP (Absolute Loss)}
                \label{fig:static_vwap_absloss_graph_\asset_\steps}
            \end{figure}
    
            \begin{figure}[H]
                \centering
                \includegraphics[width=\linewidth]
                {appendix/figures/vwap_execution_allocation_StaticVWAP_using_quadratic_vwap_loss_\asset_\steps_steps.jpg}
                \caption{StaticVWAP (Quadratic Loss)}
                \label{fig:static_vwap_quadloss_graph_\asset_\steps}
            \end{figure}
    
            \begin{figure}[H]
                \centering
                \includegraphics[width=\linewidth]
                {appendix/figures/vwap_execution_allocation_StaticVWAP_using_volume_curve_loss_\asset_\steps_steps.jpg}
                \caption{StaticVWAP (Volume Curve Loss)}
                \label{fig:static_vwap_volcurveloss_graph_\asset_\steps}
            \end{figure}
    
            \begin{figure}[H]
                \centering
                \includegraphics[width=\linewidth]
                {appendix/figures/vwap_execution_allocation_Dynamic_over_Static_VWAP_using_absolute_vwap_loss_\asset_\steps_steps.jpg}
                \caption{Dynamic StaticVWAP (Absolute Loss)}
                \label{fig:dynamic_static_vwap_absloss_graph_\asset_\steps}
            \end{figure}
    
            \begin{figure}[H]
                \centering
                \includegraphics[width=\linewidth]
                {appendix/figures/vwap_execution_allocation_Dynamic_over_Static_VWAP_using_quadratic_vwap_loss_\asset_\steps_steps.jpg}
                \caption{Dynamic StaticVWAP (Quadratic Loss)}
                \label{fig:dynamic_static_vwap_quadloss_graph_\asset_\steps}
            \end{figure}
    
            \begin{figure}[H]
                \centering
                \includegraphics[width=\linewidth]
                {appendix/figures/vwap_execution_allocation_Dynamic_over_Static_VWAP_using_volume_curve_loss_\asset_\steps_steps.jpg}
                \caption{Dynamic StaticVWAP (Volume Curve Loss)}
                \label{fig:dynamic_static_vwap_volcurveloss_graph_\asset_\steps}
            \end{figure}
        \end{multicols}
        
        \FloatBarrier
        \clearpage
    }
}

%% file: main.bbl
\begin{thebibliography}{10}

\bibitem{TotalCostOfTransactions}
S.~Berkowitz, D.~Logue, and E.~Noser.
\newblock The total cost of transactions on the nyse.
\newblock {\em Journal of Finance}, 43:97--112, 1988.

\bibitem{LeFol2006}
J.~Bialkowski, S.~Darolles, and G.~Le~Fol.
\newblock {Improving VWAP strategies: A dynamic volume approach}.
\newblock {\em {Journal of Banking and Finance}}, 32:1709--1722, 2008.

\bibitem{LeFol2012}
J.~Bialkowski, S.~Darolles, and G.~Le~Fol.
\newblock {Reducing the risk of VWAP orders execution - A new approach to modeling intra-day volume}.
\newblock {\em {JASSA}}, (1), 2012.

\bibitem{bouchard}
B.~Bouchard and N.~M. Dang.
\newblock {Generalized stochastic target problems for pricing and partial hedging under loss constraints - Application in optimal book liquidation}.
\newblock {\em {Finance and Stochastics}}, 17(1):31--72, Jan. 2013.

\bibitem{Tianhui}
R.~Carmona and T.~Li.
\newblock Dynamic programming and trade execution.
\newblock 2013.

\bibitem{Easley1987}
D.~Easley and M.~O'Hara.
\newblock Price, trade size, and information in securities markets.
\newblock {\em Journal of Financial Economics}, 19(1):69--90, 1987.

\bibitem{frei}
C.~Frei and N.~Westray.
\newblock Optimal execution of a vwap order: A stochastic control approach.
\newblock {\em Mathematical Finance}, 25, 10 2013.

\bibitem{genet2024tkat}
R.~Genet and H.~Inzirillo.
\newblock A temporal kolmogorov-arnold transformer for time series forecasting.
\newblock {\em arXiv preprint arXiv:2406.02486}, 2024.

\bibitem{genet2024tln}
R.~Genet and H.~Inzirillo.
\newblock A temporal linear network for time series forecasting.
\newblock arXiv preprint arXiv:2410.21448, 2024.

\bibitem{genet2024tkan}
R.~Genet and H.~Inzirillo.
\newblock Tkan: Temporal kolmogorov-arnold networks.
\newblock {\em arXiv preprint arXiv:2405.07344}, 2024.

\bibitem{genet2025siggate}
R.~Genet and H.~Inzirillo.
\newblock Siggate: Enhancing recurrent neural networks with signature-based gating mechanisms.
\newblock {\em arXiv preprint arXiv:2502.09318}, 2025.

\bibitem{Gourieroux}
C.~Gourieroux, J.~Jasiak, and G.~Le~Fol.
\newblock {Intra-day market activity}.
\newblock {\em Journal of Financial Markets}, 2(3):193--226, August 1999.

\bibitem{Gueant}
O.~Gu{\'e}ant and G.~Royer.
\newblock Vwap execution and guaranteed vwap.
\newblock {\em SIAM J. Financial Math.}, 5:445--471, 2014.

\bibitem{Humphery}
M.~L. Humphery-Jenner.
\newblock {Optimal VWAP trading under noisy conditions}.
\newblock {\em Journal of Banking \& Finance}, 35(9):2319--2329, September 2011.

\bibitem{inzirillo2024kamoe}
H.~Inzirillo and R.~Genet.
\newblock A gated residual kolmogorov-arnold networks for mixtures of experts.
\newblock arXiv preprint arXiv:2409.15161, 2024.

\bibitem{inzirillo2024sigkan}
H.~Inzirillo and R.~Genet.
\newblock Sigkan: Signature-weighted kolmogorov-arnold networks for time series.
\newblock {\em arXiv preprint arXiv:2406.17890}, 2024.

\bibitem{Karpoff1987}
J.~Karpoff.
\newblock The relation between price changes and trading volume: A survey.
\newblock {\em Journal of Financial and Quantitative Analysis}, 22(1):109--126, 1987.

\bibitem{kim2023adaptive}
S.~Kim, J.~Kim, H.~K. Sul, and Y.~Hong.
\newblock An adaptive dual-level reinforcement learning approach for optimal trade execution.
\newblock {\em arXiv preprint arXiv:2307.10649}, 2023.

\bibitem{Konishi}
H.~Konishi.
\newblock Optimal slice of a vwap trade.
\newblock {\em Journal of Financial Markets}, 5(2):197--221, 2002.

\bibitem{li2022hierarchical}
X.~Li, P.~Wu, C.~Zou, and Q.~Li.
\newblock Hierarchical deep reinforcement learning for vwap strategy optimization.
\newblock {\em arXiv preprint arXiv:2212.14670}, 2022.

\bibitem{Mackenzie}
M.~Mackenzie.
\newblock High frequency trading under scrutiny.
\newblock {\em Financial Times}, (1), july 2009.

\bibitem{Madhavan2002}
A.~Madhavan.
\newblock Vwap strategies, transaction performance: The changing face of trading investment guides series.
\newblock {\em Institutional Investor Inc.}, pages 32--38, 2002.

\bibitem{Culoch2007}
J.~McCulloch and V.~Kazakov.
\newblock Optimal vwap trading strategy and relative volume.
\newblock (201), 2007.

\bibitem{Culoch2012}
J.~McCulloch and V.~Kazakov.
\newblock Mean variance optimal vwap trading.
\newblock April 2012.

\bibitem{papanicolaou2023optimal}
A.~Papanicolaou, H.~Fu, P.~Krishnamurthy, B.~Healy, and F.~Khorrami.
\newblock An optimal control strategy for execution of large stock orders using lstms.
\newblock {\em arXiv preprint arXiv:2301.09705}, 2023.

\bibitem{perold1988implementation}
A.~F. Perold.
\newblock The implementation shortfall: Paper versus reality.
\newblock {\em Journal of Portfolio Management}, 14(3):4, 1988.

\bibitem{Tauchen1983}
G.~Tauchen and M.~Pitts.
\newblock The price variability-volume relationship on speculative markets.
\newblock {\em Econometrica}, 51(2):485--505, 1983.

\bibitem{Foster}
S.~Viswanathan and F.~Foster.
\newblock A theory of the interday variations in volume, variance, and trading costs in securities markets.
\newblock {\em Review of Financial Studies}, 3:593--624, 02 1990.

\end{thebibliography}
